\documentclass[twocolumn,aps,superscriptaddress,longbibliography]{revtex4-1}
\usepackage{bm}%
\usepackage{float}
\expandafter\ifx\csname package@font\endcsname\relax\else
 \expandafter\expandafter
 \expandafter\usepackage
 \expandafter\expandafter
 \expandafter{\csname package@font\endcsname}%
\fi

\usepackage{array}
\usepackage{amsmath,amssymb}
\usepackage{MnSymbol}
 \usepackage{tikz-feynman} 

\usepackage{graphicx}
\usepackage{mathrsfs}
\usepackage{bm}
\usepackage{mathtools}
\usepackage{epstopdf}
\usepackage{hyperref}

\hypersetup{%
   pdfpagemode=None, %FullScreen,
   pdfstartpage=1,
   pdfmenubar=true,
   pdftoolbar=true,
   colorlinks = true,
   linkcolor=blue,
   citecolor=blue,
   urlcolor=blue,
   bookmarksopen=false
 }
 \usepackage{amsmath}
\usepackage{amsfonts}
\usepackage{mathtools}
\usepackage{graphicx}
\usepackage{fixme}
\usepackage{color}
\usepackage{soul}

\begin{document}
\title{Strong interactions and biexcitons in a polariton mixture}
\author{M. A. Bastarrachea-Magnani, A. Camacho-Guardian}
\affiliation{ 
Department of Physics and Astronomy, Aarhus University, Ny Munkegade, DK-8000 Aarhus C, Denmark. }
\author{M. Wouters}
\affiliation{TQC, Universiteit Antwerpen, Antwerpen, B-2610, Belgium.}
\author{G. M.\ Bruun} 
\affiliation{ 
Department of Physics and Astronomy, Aarhus University, Ny Munkegade, DK-8000 Aarhus C, Denmark. }
\affiliation{Shenzhen Institute for Quantum Science and Engineering and Department of Physics, Southern University of Science and Technology, Shenzhen 518055, China}
\begin{abstract}
We develop a many-body theory for the properties of  exciton-polaritons interacting strongly with a Bose-Einstein condensate (BEC) of 
exciton-polaritons in another spin state. Interactions lead to   the presence of  a two-body bound state, the biexciton, giving rise to a Feshbach resonance in the polariton spectrum when its energy is equal to that of two free polaritons. Using the minimal set of terms to describe this resonance, our theory recovers the main findings of two experiments probing interaction effects for upper and lower polaritons in a BEC. This strongly supports that Feshbach physics has indeed been realized, and we furthermore extract the energy and decay of the biexciton from the experimental data. The decay rate is predicted to be much larger than that coming from its dissociation into two free polaritons indicating that other decay channels are important. 
\end{abstract} 
\maketitle

%%%%%%%%%%%%%%%%%%%%%%%%%%%%%%%%%%%%%%%%%%%%%%%%%%
%%%%%%%%%%%%%%%%%%%%%%%%%%%%%%%%%%%%%%%%%%%%%%%%%%
%%%%%%%%%%%%%%%%%%%%%%%%%%%%%%%%%%%%%%%%%%%%%%%%%%

\section{Introduction}
The strong coupling between cavity photons and excitons in semiconductor microcavities gives rise to the formation of  quasiparticles, the exciton-polaritons~\cite{Hopfield1958,Weisbuch1992,Carusotto2013,Kavokin2017}. Exciton-polaritons (polaritons from now on) mix properties of light and matter and their study has produced several breakthrough results such as the observation of Bose-Einstein condensation and superfluidity~\cite{Kasprzak2006,Wouters2007a,Amo2009,Deng2010,Kohnle2011,Kohnle2012}, quantum vortices~\cite{Lagoudakis2008}, and topological states of light~\cite{St-Jean2017,Klembt2018}. A promising research direction with great technological potential is to exploit the Coulomb interaction between the excitonic part of polaritons to induce strong interactions between their photonic part~\cite{Sanvitto2016,Munoz-Matutano2019,Delteil2019}. The interaction  can be enhanced when the electrons are in a strongly correlated  state~\cite{Knuppel2019} or by using dipolar excitons~\cite{Cristofolini2012,Byrnes2014,Rosenberg2018,Togan2018}. Recently, strong interactions have been observed using polarization-resolved pump-probe spectroscopy to create a mixture of polaritons in two spins states, with one component forming a Bose-Einstein condensate (BEC)~\cite{Takemura2014,Takemura2014b,Takemura2017,Navadeh2019}. Large energy shifts were observed, which were attributed to a Feshbach resonance mediated by a biexciton state by means of a mean-field two-channel theory and a non-equilibrium  Gross-Pitaevskii equation~\cite{LaRocca1998,Borri2000,Borri2003,Ivanov2004}. This opens up the exciting possibility to control the strength and sign of inter-excitonic interaction in analogy with the case of  degenerate atomic gases,  where Feshbach resonances have  become a very powerful tool~\cite{Chin2010,Kokkelmans2014}.

Here, we present a  strong coupling theory describing the formation of biexcitons in a spin mixture of polaritons created in  pump-probe experiments~\cite{Takemura2014,Takemura2014b,Takemura2017,Navadeh2019}. Our theory is minimal in the sense that it contains precisely the terms necessary to describe Feshbach  interactions  due to the formation of a  biexciton. We show that it recovers the main  results of the two recent experiments probing energy shifts in polariton mixtures~\cite{Takemura2017,Navadeh2019}, supporting the finding that a Feshbach mediated interaction was indeed observed. Moreover, we extract both the energy and decay rate of the biexciton state from the experimental data, and predict that the latter is much larger than can be explained from the splitting of the biexciton into two unbound polaritons. 

%\emph{System.--}
\section{System and Hamiltonian}

We consider a two-dimensional (2D) mixture of polaritons in spin states $\sigma=\{\downarrow, \uparrow\}$, resulting from the coupling of two quantum light fields with counter-circular polarizations to the excitonic modes of a 2D semiconductor microcavity~\cite{Takemura2014,Takemura2014b,Takemura2016,Takemura2017,Navadeh2019}. The Hamiltonian  is 
\begin{align}
\hat{H}=\sum_{\mathbf{k}\sigma}
\begin{bmatrix}\hat x_{\mathbf{k}\sigma}^\dagger&\hat c_{\mathbf{k}\sigma}^\dagger\end{bmatrix}
\begin{bmatrix}
\varepsilon_{\mathbf{k}}^{x} & \Omega \\
\Omega & \varepsilon_{\mathbf{k}}^{c}
\end{bmatrix}
\begin{bmatrix}
\hat x_{\mathbf{k}\sigma} \\
\hat c_{\mathbf{k}\sigma}
\end{bmatrix}\nonumber\\
+\frac{g}{2}\sum_{\mathbf{q},\mathbf{k},\mathbf{k}'}\hat{x}_{\mathbf{k}+\mathbf{q}\uparrow}^{\dagger}\hat{x}_{\mathbf{k}'-\mathbf{q}\downarrow}^{\dagger}
\hat{x}_{\mathbf{k}'\downarrow}\hat{x}_{\mathbf{k}\uparrow}.
\label{eq:Hamiltonian}
\end{align}
Here, $\hat{x}^{\dagger}_{\mathbf{k}\sigma}$ and $\hat{c}^{\dagger}_{\mathbf{k}\sigma}$ create an exciton and a photon respectively with momentum  $\mathbf{k}$ and spin $\sigma$. These states have the kinetic energies $\varepsilon_{\mathbf{k}}^{x}=\mathbf{k}^{2}/2m_{x}$ and $\varepsilon_{\mathbf{k}}^{c}=\mathbf{k}^{2}/2m_{c}+\delta$, where $m_x$ and $m_c$ are the masses of the exciton and the cavity photon, and we use units where the system volume, $\hbar$, $c$, and $k_B$ are all one. The detuning $\delta$  at zero momentum can be controlled  experimentally, and  $\Omega$ gives the  strength of the spin-conserving photon-exciton coupling, which is taken to be real. Excitons with opposite spin interact with  the strength $g$, which 
is momentum independent to a good approximation, since its  typical length scale  is set by the exciton radius, which is much shorter than the other length scales in the problem. Since the aim of this paper is to explain experimental results using the simplest model possible, we ignore terms describing the interaction between excitons with parallel spin as well as a  photon-assisted scattering term~\cite{Ciuti2003,Combescot2007,Ciuti2005}. 

We consider experiments where the concentration of  $\uparrow$ polaritons created by a pump beam, is much larger than that of $\downarrow$ polaritons created by a probe beam with lower intensity~\cite{Takemura2017,Navadeh2019}. It follows that we can regard the $\uparrow$ polaritons as unaffected by the presence of the $\downarrow$ polaritons. Their dispersion is then easily obtained by diagonalizing the Hamiltonian in Eq.~\eqref{eq:Hamiltonian} without the interaction term $\hat H_\text{int}$, giving the well-known lower- and upper-polaritons
\begin{align}
\begin{bmatrix}
\hat{x}_{\mathbf{k}\uparrow} \\
\hat{c}_{\mathbf{k}\uparrow}
\end{bmatrix}
=
\begin{bmatrix}
\mathcal{C}_{\mathbf{k}} & -\mathcal{S}_{\mathbf{k}} \\
\mathcal{S}_{\mathbf{k}} & \mathcal{C}_{\mathbf{k}} \\
\end{bmatrix}
\begin{bmatrix}
\hat{L}_{\uparrow,\mathbf{k}} \\
\hat{U}_{\uparrow,\mathbf{k}}
\end{bmatrix}
\label{eq:Polaritons}
\end{align}
created by the $\hat{L}_{\uparrow,\mathbf{k}}^\dagger$ and $\hat{U}_{\uparrow,\mathbf{k}}^{\dagger}$ operators with  energies %are
 \begin{align} \label{eq:Polaritonenergies}
\varepsilon^\text{UP,LP}_{\mathbf{k}}=\frac{1}{2}\left(\delta_{\mathbf{k}}+2\varepsilon_{\mathbf{k}}^{x}\pm\sqrt{\delta_{\mathbf{k}}^{2}+4\Omega^{2}}\right).
\end{align}
The corresponding Hopfield coefficients are~\cite{Hopfield1958}
\begin{align}
\mathcal{C}^{2}_{\mathbf{k}}=\frac{1}{2}\left(1+\frac{\delta_{\mathbf{k}}}{\sqrt{\delta_{\mathbf{k}}^{2}+4\Omega^{2}}}\right) \text{ and } 
\mathcal{S}^{2}_{\mathbf{k}}=1-\mathcal{C}^{2}_{\mathbf{k}}
\end{align}
where $\delta_{\mathbf{k}}=\varepsilon_{\mathbf{k}}^{c}-\varepsilon_{\mathbf{k}}^{x}$ is the momentum-dependent detuning. 
In accordance with the experiments, we assume that the pump beam creates a BEC of $\uparrow$ lower polaritons in the ${\mathbf k}=0$ state with density $n_{\uparrow}^\text{LP}$~\cite{Takemura2017,Navadeh2019}. 

%\emph{Diagrammatic approach.--} 
\section{Diagrammatic approach}

Our focus is on Feshbach mediated interaction effects on the $\downarrow$ polaritons. Despite the fact that polariton systems are driven by external lasers, many of their steady-state properties can be accurately described using equilibrium theory with a few modifications, such as the chemical potentials being determined by the external laser frequencies~\cite{Carusotto2013}. We  therefore consider the finite temperature Green's function ${\mathcal G}_\downarrow(\mathbf k,\tau)=-\langle T_\tau \{\hat\Psi_{\mathbf k}(\tau)\hat\Psi_{\mathbf k}^\dagger(0)\}\rangle$ where $\hat\Psi_{\mathbf k}=[\hat x_{\mathbf k\downarrow},\hat c_{\mathbf k\downarrow}]^T$ and $T_\tau$ denotes the imaginary time ordering. We have 
\begin{align}
{\mathcal G}^{-1}_\downarrow(\mathbf k,i\omega_n)=
\begin{bmatrix}
i\omega_n-\varepsilon_{\mathbf{k}}^{x}& \Omega \\
\Omega^* & i\omega_n-\varepsilon_{\mathbf{k}}^{c}
\end{bmatrix}
-
\begin{bmatrix}
\Sigma({\mathbf k},i\omega_n)& 0 \\
0 & 0
\end{bmatrix}
\label{eq:GreensFn}
\end{align} 
where $\omega_n=2nT$ with $n=0,\pm1,\ldots$ is a bosonic Matsubara frequency, $T$ is the temperature, and interaction effects are described by the exciton self-energy $\Sigma$. Given their low concentration, the $\downarrow$ excitons can be regarded as mobile impurities in the BEC of $\uparrow$ polaritons. 

\begin{figure}[!ht]
\begin{center}
\includegraphics[width=0.45\textwidth]{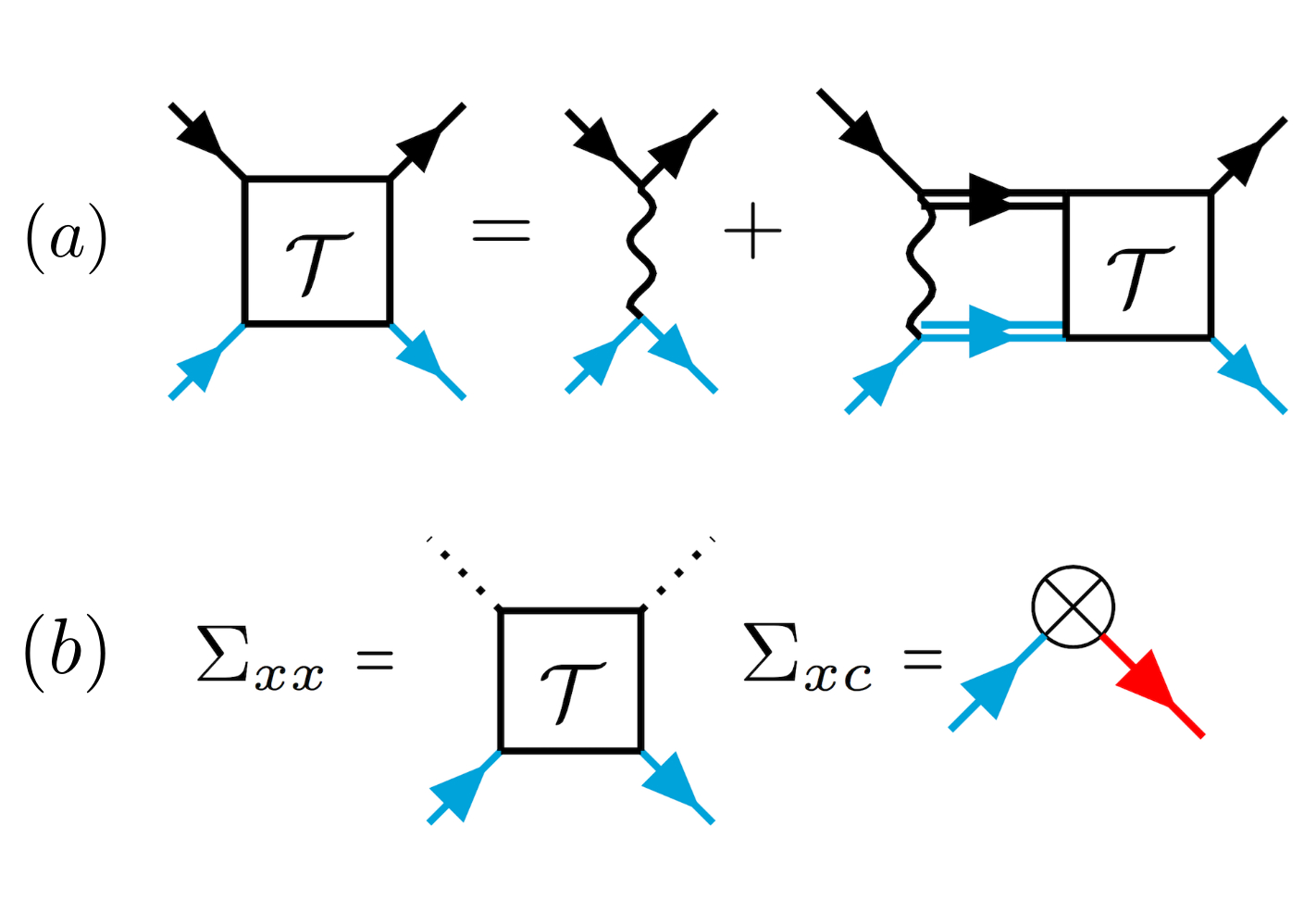}
\end{center}
\caption{(Color online). (a) The $\uparrow$-exciton $\downarrow$-exciton scattering matrix in the ladder approximation including the light-matter coupling. Spin $\uparrow$- and $\downarrow$-excitons are given by black and cyan lines, and double lines indicate that they are expressed in the polariton basis. The wiggly line is the 
$\uparrow\downarrow$-interaction. (b) The exciton self-energy. The crossed circle is the light-matter coupling $\Omega$, and cavity photon and condensate $\uparrow$-excitons are denoted by red  and black dashed lines respectively.}
\label{fig:Feynman} 
\end{figure}

We calculate $\Sigma$ using a ladder approximation illustrated in Fig.~\ref{fig:Feynman}, which includes the two-body Feshbach correlations exactly. 
Such an approach has turned out to be surprisingly accurate for impurities in atomic gases~\cite{Massignan2014,Rath2013,Ardila2019}, and  here we 
generalise it to include strong light coupling. The self-energy becomes  
\begin{align}
\Sigma({\mathbf k},i\omega_n)=n_{x\uparrow}\mathcal{T}({\mathbf k},i\omega_n),
\end{align} 
describing the scattering of  an $\uparrow$-exciton out of the BEC by the $\downarrow$-exciton. 
Here, $n_{x\uparrow}=\mathcal{C}^{2}_{\mathbf{k}=0}n_{\text{LP}\uparrow}$ is the density of the excitonic component  of the 
$\uparrow$ polariton BEC, and 
\begin{align}
\mathcal{T}({\mathbf k},i\omega_n)=\frac1{\text{Re}\Pi_V(E_B)-\Pi({\mathbf k},i\omega_n)+i\gamma} 
\label{eq:Tmatrix}
\end{align} 
is the scattering matrix with 
\begin{gather}\label{eq:Pair}
\Pi({\mathbf k},i\omega_n)=-T\sum_{{\mathbf q},i\omega_l}\mathcal{G}_{x}^{\downarrow}({\mathbf k}-\mathbf{q},i\omega_n-i\omega_l )
\mathcal{G}_{x}^{\uparrow}(\mathbf{q},i\omega_l)
\end{gather}
being the propagator of a pair of excitons with center-of-mass momentum $\mathbf{k}$.
  As detailed in the Appendix~\ref{app:1},  the coupling strength $g$ is eliminated in favor of the zero momentum exciton pair propagator in a vacuum $\Pi_V$, evaluated at the binding energy $E_B$ of a $\uparrow$- and $\downarrow$-exciton pair~\cite{Randeria1990,Wouters2007,Carusotto2010}. In this way, the biexciton state giving rise to Feshbach physics naturally emerges from our formalism as a pole in the $\mathcal{T}$ matrix. We have also introduced a phenomenological decay $\gamma$, as discussed in detail below. 
  %%%%%%%%%%%%%%%%%%%%%%%%%%%%%%%%%%%%%%%%%%%%%%%%%%
%%%%%%%%%%%%%%%%%%%%%%%%%%%%%%%%%%%%%%%%%%%%%%%%%%

Since the coupling $\Omega$ to light is strong, we express the exciton propagators in Eq.~\eqref{eq:Pair}  in the polariton basis as 
\begin{align}
\mathcal{G}_{x}^{\downarrow}(\mathbf{k},i\omega_{n})&=\frac{\mathcal{C}^{2}_{\mathbf{k}}}{i\omega_{n}-\varepsilon_{\mathbf{k}\downarrow}^{LP}}+\frac{\mathcal{S}^{2}_{\mathbf{k}}}{i\omega_{n}-\varepsilon_{\mathbf{k}\downarrow}^{UP}},\nonumber\\
\mathcal{G}_{x}^{\uparrow}(\mathbf{k},i\omega_{n})&=\frac{\mathcal{C}^{2}_{\mathbf{k}}}{i\omega_{n}-\varepsilon_{\mathbf{k}\uparrow}^{LP}+\varepsilon^{LP}_{0}}+\frac{\mathcal{S}^{2}_{\mathbf{k}}}{i\omega_{n}-\varepsilon_{\mathbf{k}\uparrow}^{UP}+\varepsilon^{LP}_{0}}.
\label{eq:Polaritonbasis}
\end{align} 
Here, $\varepsilon^\text{UP,LP}_{\mathbf{k}\downarrow}=\varepsilon^\text{UP,LP}_\mathbf{k}$, whereas $\varepsilon^\text{UP,LP}_{\mathbf{k}\uparrow}=\varepsilon^\text{UP,LP}
_\mathbf{k}-\varepsilon_{0}^{LP}$ since the energy of the $\uparrow$ polaritons is measured with respect to the  chemical potential of the BEC, which we take to be ideal.
The Matsubara sum in Eq.~\eqref{eq:Pair} can now easily be performed, which for  zero temperature gives 
\begin{gather}
\Pi(k)=\int\!\frac{{d^{2}\mathbf{q}}}{(2\pi)^{2}}\left(
\frac{\mathcal{C}^{2}_{\mathbf{k}+\mathbf{q}}\mathcal{C}^{2}_{\mathbf{q}}}{i\omega_{n}-\varepsilon^{LP}_{\mathbf{k}+\mathbf{q}\uparrow}-\varepsilon^{LP}_{\mathbf{q}\downarrow}}+\frac{\mathcal{C}^{2}_{\mathbf{k}+\mathbf{q}}\mathcal{S}^{2}_{\mathbf{q}}}{i\omega_{n}-\varepsilon^{LP}_{\mathbf{k}+\mathbf{q}\uparrow}-\varepsilon^{UP}_{\mathbf{q}\downarrow}}\right.\nonumber\\ 
+ \left.\frac{\mathcal{S}^{2}_{\mathbf{k}+\mathbf{q}}\mathcal{C}^{2}_{\mathbf{q}}}{i\omega_{n}-\varepsilon^{UP}_{\mathbf{k}+\mathbf{q}\uparrow}-\varepsilon^{LP}_{\mathbf{q}\downarrow}}+\frac{\mathcal{S}^{2}_{\mathbf{k}+\mathbf{q}}\mathcal{S}^{2}_{\mathbf{q}}}{i\omega_{n}-\varepsilon^{UP}_{\mathbf{k}+\mathbf{q}\uparrow}-\varepsilon^{UP}_{\mathbf{q}\downarrow}}\right)
\label{eq:Excitonintermsofpolaritons}
\end{gather} 
where $k\equiv({\mathbf k},i\omega_n)$.
 Equation~\eqref{eq:Excitonintermsofpolaritons} is a first iteration in a self-consistent calculation, in the sense that  the propagators obtained by diagonalizing the first term in Eq.~\eqref{eq:GreensFn} are used. A similar approach was recently used in the context of Fermi polaron-polaritons~\cite{Tan2019}.

Finally, the retarded $\downarrow$ Green's function is obtained by the usual analytic continuation $G_\downarrow(\mathbf k,\omega)=\left.{\mathcal G}_\downarrow(\mathbf k,i\omega_n)\right|_{i\omega_n\rightarrow \omega+i0_+}$~\cite{Fetter1971}. Further details on the diagrammatic scheme are given in the Appendices.

%{\em biexciton energy and decay.--} 
\section{biexciton energy and decay}

We now discuss the energy and decay of the biexciton state. The effective propagator for the biexciton is obtained from a pole expansion of the $\mathcal{T}$ matrix in Eq.~\eqref{eq:Tmatrix} giving 
\begin{align}
\mathcal{T}(\mathbf{k},\omega)\simeq\frac{g_{eff}}{\omega-E_B(\mathbf k)+\varepsilon^\text{LP}_{0}+ig_{eff}\mbox{Im}\left[\Pi(\mathbf{k},\omega)\right]+i\gamma_B}
\label{eq:Poleexpansion}
\end{align}
with coupling between the excitons and the biexciton
\begin{gather}
g_\text{eff}=-\left.\left\{\frac{\partial}{\partial\omega}\mbox{Re}[\Pi(\mathbf{k},\omega)]\right\}^{-1}\right|_{E_B({\mathbf k})}. 
\end{gather}
Here, $E_B(\mathbf k)-\varepsilon^\text{LP}_{0}\simeq E_B+k^2/2(m_x+m_c)-\varepsilon^\text{LP}_{0}$ is the 
the pole of the $\mathcal{T}$-matrix determining the biexciton energy with momentum ${\mathbf k}$ in the presence of light. There are two contributions to the biexciton decay: $\gamma_{B}^{rad}=g_{eff}\mbox{Im}\left[\Pi(\mathbf{k},\omega)\right]$ describes the dissociation of the biexciton into two free polaritons, and $\gamma_B=g_\text{eff}\gamma$ describes an additional decay, for instance due to disorder. 
 
%\emph{Comparison with experiment.--} 
The goal is to analyze two different experiments using our minimal model:
In Ref.~\cite{Navadeh2019}, the energy of an \emph{upper} $\downarrow$ polariton in a BEC of lower $\uparrow$ polaritons was measured, and the same was done for a \emph{lower} $\downarrow$ polariton  in Ref.~\cite{Takemura2017}. To model these experiments, we have used the typical values for GaAs-based microcavities $2\Omega\simeq 3.5m\mbox{eV}$~\cite{Borri2003,Takemura2014}, $m_{x}=0.25\,m_{e}$, where $m_{e}$ is the free electron mass, and $m_{c}=10^{-4}m_{x}$~\cite{Alessi2000,Yamamoto2000,Deveaud2012}. There is a substantial experimental uncertainty concerning the density $n_{\text{LP}\uparrow}$ of the polariton BEC, which was not measured directly. We take $n_{\text{LP}\uparrow}=\mathcal{S}_{0}^{2}n_{pu}$, where $n_{pu}$ is proportional to the density of the photons in the pump beam, and our numerical results are calculated for $n_{pu}\simeq 3.7\mbox{x}10^{10} \mbox{cm}^{-2}$\footnote{The reported value of the density $n_{pu}$ in $\mbox{cm}^{-2}$ depends on the chosen value of the exciton mass. As typical values range from $m_{x}=0.1m_{e}$ to $0.25m_{e}$ we could have an uncertainty of 50\% of the chosen value for the density.}. Additionally, we introduce a small lower-polariton decay at $\gamma_{0}^{LP}\sim0.1m\mbox{eV}$ and a small  broadening of the impurity propagator $\eta=0.07m\mbox{eV}$ in order to numerically resolve the polariton energy. 

%%%%%%%%%%%%%%%%%%%%%%%%%%%%%%%%%%%%%%%%%%%%%%%%%%%%%%%%%%%%%%%%%%
\begin{figure}%[!ht]
\begin{center}
\includegraphics[width=0.48\textwidth]{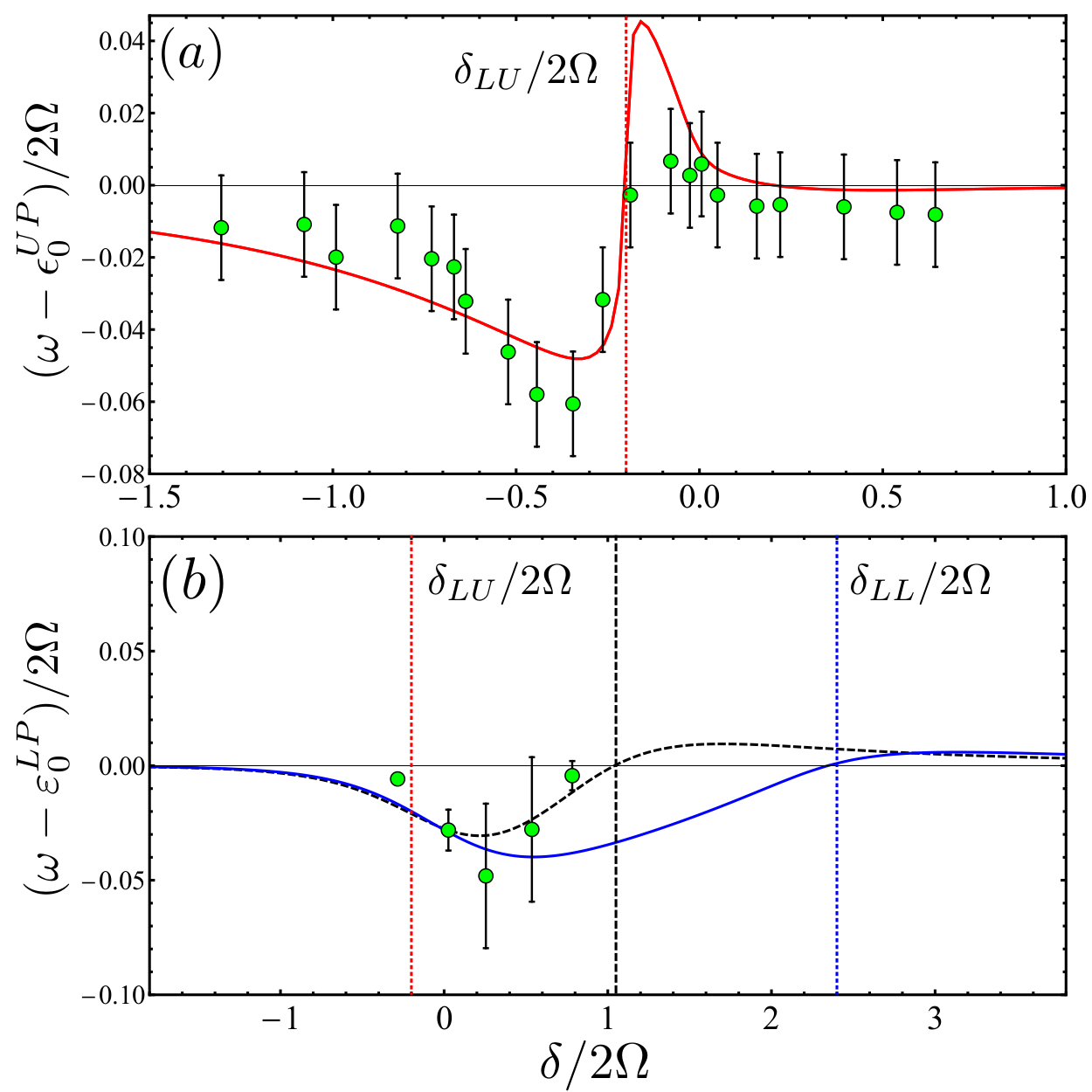}
\end{center}
\vspace{-15pt}
\caption{(Color online). The energy of the (a) upper and (b) lower $\downarrow$ polariton in a BEC of $\uparrow$ polaritons as a function of the detuning for $E_{B}=-0.7\mbox{meV}$ and $\gamma_{B}=0.4\mbox{meV}$, along with experimental data from Refs.~\cite{Takemura2017,Navadeh2019} (green points). The biexciton energy $E_B$ equals that of two upper-lower or two lower-lower polaritons at the red and blue vertical dotted lines. The black dashed curve in panel (b) corresponds to $E_{B}=-1.4\mbox{meV}$ and $\gamma_{B}=1.28\mbox{meV}$ with a corresponding black dashed vertical line where the biexciton energy equals that of two lower polaritons.}
\label{fig:1} 
\end{figure} 
%%%%%%%%%%%%%%%%%%%%%%%%%%%%%%%%%%%%%%%%%%%%%%%%%%%%%%%%%%%%%%%%%%

Figure~\ref{fig:1} shows our main numerical results. In Fig.~\ref{fig:1} (a), we plot the energy shift of the upper $\downarrow$ polariton due to interactions with the BEC as a function of the detuning $\delta$, alongside the experimental results of Ref.~\cite{Navadeh2019}. There are two free parameters  in our theory: the binding energy $E_{B}$ and the decay $\gamma_B$ of the biexciton, which determine the position and width of the resonance. The dashed vertical line in Fig.~\ref{fig:1} (a) indicates the value of the detuning $\delta_{LU}=E_{B}$ where the energy of the biexciton equals that of an upper $\downarrow$ polariton plus a lower $\uparrow$ polariton, i.e.\ $E_{B}=\varepsilon^{LP}_{0}+\varepsilon^{UP}_{0}$. For detunings close to $\delta_{LU}$, the upper $\downarrow$ polariton interacts strongly with the BEC via the formation of the biexciton. The result is a resonance feature broadened by the biexciton decay. We have chosen $E_{B}=-0.7\mbox{meV}$ and $\gamma_{B}=0.4\mbox{meV}$, for which the theory can be seen to agree well with the experimental results. The binding energy is close to experimental values~\cite{Borri2000,Navadeh2019}, and below the upper limit  of $-1.4\mbox{meV}$ for an infinitely deep confinement potential~\cite{Birkedal1996,Navadeh2019}.

In Fig.~\ref{fig:1} (b), we show as a solid blue line the energy of the lower $\downarrow$ polariton with respect to its non-interacting value, as a function of the detuning $\delta$, along with the experimental results of Ref.~\cite{Takemura2017}. There is reasonable agreement between theory and experiment, in particular concerning the position as well as the depth of the resonance feature. The position is determined by the energy crossing of the lower $\uparrow$ and $\downarrow$ polaritons with the biexciton state, i.e., $E_{B}=2\varepsilon_{0}^{LP}$ giving $\delta_{LL}=(E_{B}^{2}-4\Omega^{2})/2E_{B}$ shown by the dashed vertical line. We emphasize that the agreement is obtained with no fitting, as we use the values $E_{B}=-0.7\mbox{meV}$ and $\gamma_{B}=0.4\mbox{meV}$ extracted from the fit to the other experiment described above. 

We conclude that our strong coupling theory agrees well with two experiments measuring the energy of an upper and a lower $\downarrow$ polariton in a BEC of lower $\uparrow$ polaritons. Importantly, this agreement is achieved using the \emph{same} values for the biexciton energy and decay for the two experiments, which makes physically sense since it is the same biexciton that mediates the Feshbach interaction in the two experiments. An even better fit can surely be achieved using a more extensive numerical fitting to both experiments simultaneously or by introducing more terms in our theory such as interactions between parallel spin excitons. Such an approach is however not supported by the experimental data, which are characterised by  significant uncertainties, in particular concerning the BEC density. For instance,  an equally good fit can be obtained by changing the density $n_{pu}$ and the biexciton decay by up to a factor of two. On the contrary, the biexciton energy giving the position of the resonances is determined to within $\sim20\%$ by the data. The agreement using a theory with the minimal set of terms  strongly supports that these experiments indeed are indeed observing Feshbach physics, and it moreover allows us to determine the biexciton energy fairly accurately. 

We can of course also obtain even better agreement with the experimental results Ref.~\cite{Takemura2017} if we use different values of $E_B$ and $\gamma_B$, although this makes little physical sense. To illustrate this, we  plot as a dashed black curve in Fig.~\ref{fig:1} (b) the theoretical prediction for the lower $\downarrow$ polariton using $E_{B}=-1.4\mbox{meV}$  and $\gamma_{B}=1.28\mbox{meV}$, which agrees well with the experimental results. Previous theoretical analysis was based on a mean-field theory with many free parameters in addition to the binding energy and decay of the biexciton, which were taken to be different in order to fit the two experiments.

%\emph{Excitation spectra.--} 
\section{Excitation spectra}

We now provide further details of the Feshbach resonance effects observed in Fig.~\ref{fig:1}.
%%%%%%%%%%%%%%%%%%%%%%%%%%%%%%%%%%%%%%%%%%%%%%%%%%%%%%%%%%%%%%%% 
\begin{figure}
\begin{center}
\includegraphics[width=0.5\textwidth]{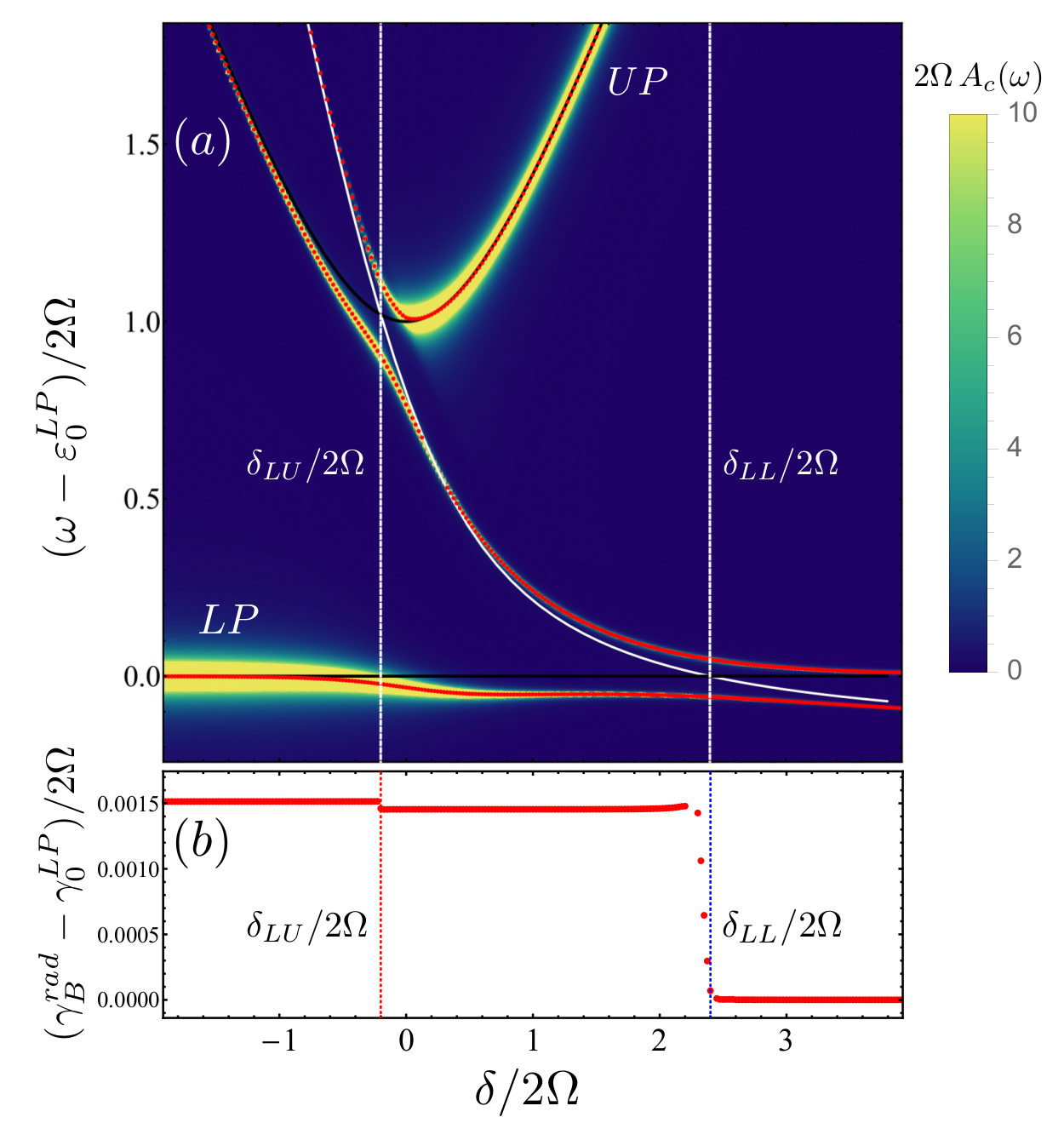} 
\end{center}
\vspace{-15pt}
\caption{(Color online). (a) The photon spectral function $A_{c}(\omega)$ as a function of the energy $\omega$ and the detuning $\delta$ for $E_{B}=-0.7\mbox{meV}$ and $\gamma_{B}=0$. The quasiparticle energies are indicated with red points, the non-interacting polariton energies are indicated with black curves, and the biexciton energy is shown with the white curve. The biexciton energy $E_B$ equals that of two upper-lower or two lower-lower polaritons at the  vertical dashed lines. (b) The damping of the biexciton due to its splitting into two free polaritons.}
\label{fig:2} 
\end{figure} 
%%%%%%%%%%%%%%%%%%%%%%%%%%%%%%%%%%%%%%%%%%%%%%%%%%%%%%%%%%%%%%%% 
In Fig.~\ref{fig:2} (a), we show the $\downarrow$ photonic spectral density $A_{c}(\omega)=-2\mbox{Im}\left[G_{cc}(\mathbf{k}=0,\omega)\right]$ giving the excitation spectrum of the $\downarrow$ polaritons as a function of the detuning and the energy. We also plot $E_B-\varepsilon^\text{LP}_{\mathbf{k}=0}$, which is the energy needed for the $\downarrow$ polariton to form a biexciton with a $\uparrow$ lower polariton from the BEC. We have used the same parameters as for Fig.~\ref{fig:1} except there is no additional damping to the biexciton, i.e.\ $\gamma_{B}=0$. CComparing the resulting polariton energies to those in the absence of interactions plotted as solid black lines from  Eq.~\eqref{eq:Polaritonenergies}, we see that interactions have significant effects.The upper-polariton branch is split into two branches with comparable spectral weight around the detuning $\delta_{LU}$ where the biexciton energy equals that of an upper and lower polariton, indicated by the left vertical line in Fig.~\ref{fig:2}. As a result, there are three branches instead of two for detunings close to $\delta_{LU}$. This splitting is  caused by a cross-polaritonic Feshbach resonance, where a $\downarrow$ upper-polariton forms a biexciton with the a $\uparrow$ lower-polariton. A similar splitting occurs around $\delta_{LL}$, where the energy of the biexciton equals that of two lower-polaritons, leading to a resonant interaction. The spectral weight of this splitting is small, reflecting that most of the photonic spectral weight is located in the upper polariton branch for positive detunings. The excitonic spectral function, not shown for brevity, is complementary  having a large spectral weight in the lower polariton branch for positive detuning. Similar results were recently presented  based on a Chevy type variational wave function including trimer states~\cite{Levinsen2018}. Note that excited exciton states can play a role for high energy (blue detuning) as was recently 
observed in a perovskite crystal~\cite{bao2018}. However, these high energies are not relevant for determining the resonance properties, which is the focus of the present 
paper. 
   
While Fig.~\ref{fig:2} qualitatively explains the experimental results in Fig.~\ref{fig:1}, the resonance features are too 
sharp. The reason is that setting $\gamma_B=0$ strongly underestimates the biexciton decay rate. To illustrate this, we plot in Fig.~\ref{fig:2} (b) the decay rate of the biexciton into two free polaritons, which from Eq.~\eqref{eq:Poleexpansion} is given by $\gamma^{rad}_{B}=g_{eff}\mbox{Im}[\Pi(\mathbf{k}=0,E_B)]$. The decay rate is approximately constant at $\sim5\mu\mbox{eV}$ for $\delta<\delta_{LL}$, confirming previous results~\cite{Carusotto2010}. This value is determined by the 2D density of states of the polariton pairs, and it is approximately two orders of magnitude too small compared to the value $\gamma_{B}=0.4\mbox{meV}$ needed to explain the experiments. 
Even though the damping rate $\gamma_B$ depends on the BEC density, which is poorly determined experimentally, this conclusion is robust, since any reasonable variation of 
the density
cannot bridge a gap of two orders of magnitude.  
For detunings $\delta>\delta_{LL}$, the biexciton energy is below the continuum of states formed by $\uparrow$ and $\downarrow$ lower-polariton pairs and the damping is zero. The decay rate also exhibits an abrupt decrease when biexciton energy falls below the continuum of states formed by an $\uparrow$ lower-polariton and an $\downarrow$ upper-polariton at $\delta_{LU}$. But the decrease is very small as can be seen from Fig.~\ref{fig:1}, since the Hopfield coefficients strongly suppress this decay channel. Note that in order to calculate the biexciton decay rate due to its dissociation into polaritons, it is crucial to express the Green's functions inside the pair propagator in the polariton basis as in Eq.~\eqref{eq:Excitonintermsofpolaritons}. A standard ladder approximation using bare exciton propagators or 
an equivalent variational calculation will not capture the decay rate due to dissociation correctly. 

\begin{figure}[!ht]
\begin{center}
\includegraphics[width=0.5\textwidth]{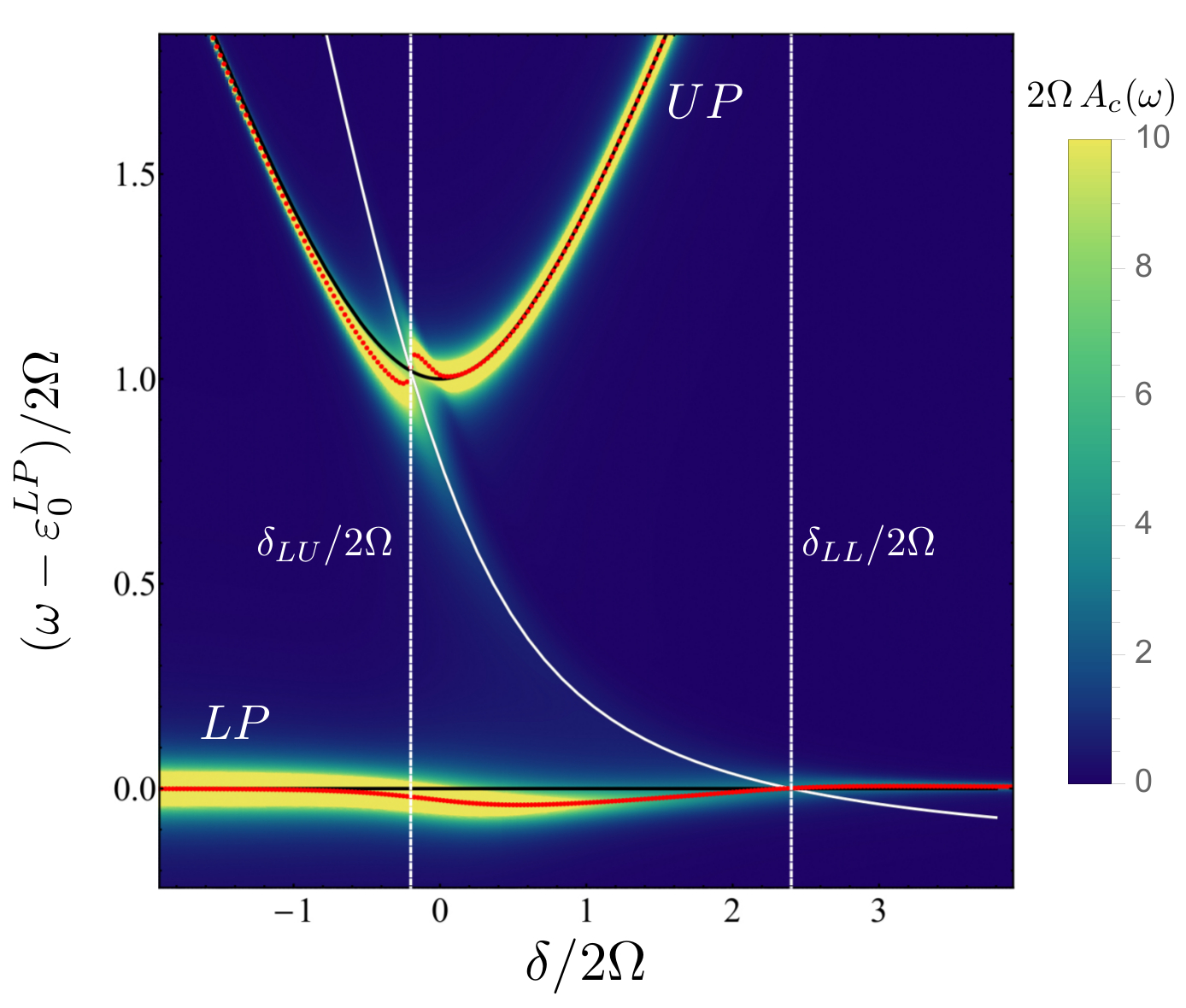}
\end{center}
\vspace{-15pt}
\caption{(Color online). Same as in Fig.~\ref{fig:2} (a), but now with $E_{B}=-0.7\mbox{meV}$ and $\gamma_{B}=0.4\mbox{meV}$.}
\label{fig:3} 
\end{figure}

In Fig.~\ref{fig:3} we plot the $\downarrow$ photonic spectral density $A_{c}(\omega)$ for the same parameters as above but now including the additional decay $\gamma_{B}=0.4\mbox{meV}$. 
%%%%%%%%%%%%%%%%%%%%%%%%%%%%%%%%%%%%%%%%%%%%%%%%%%%%%%%%%%%%%%%% 
%%%%%%%%%%%%%%%%%%%%%%%%%%%%%%%%%%%%%%%%%%%%%%%%%%%%%%%%%%%%%%%% 
Comparing Fig.~\ref{fig:3} with Fig.~\ref{fig:2}, the main effect of the added decay rate is that instead of producing a splitting of the quasiparticle branches, there are now only two branches for all detunings, the upper- and lower- polaritons. They, however, experience strong energy shifts away from the non-interacting values (black lines) at the detunings $\delta_{LU}$ and $\delta_{LL}$, respectively, which correspond directly to the shifts plotted in Fig.~\ref{fig:1}. From this, we conclude that in order to explain the experimental findings, it is essential to include a strong decay channel of the biexciton state in addition to its splitting into polariton pairs. We speculate that this additional decay could be due to disorder.
 
%{\em Discussion and conclusion.--} 
\section{Conclusions}

We presented a non-perturbative theory that describes polaritons in a BEC of polaritons in another spin state, as created  
in recent pump-probe experiments using semiconductor microcavities~\cite{Takemura2017,Navadeh2019}. Our theory contains the minimal set of terms describing interactions mediated via the formation of a biexciton, and it recovers the main findings of  the two experiments probing energy shifts of the upper and lower polaritons. In addition to confirming that strong interactions via a Feshbach resonance were indeed realized, this also enabled us to extract the energy and decay rate of the biexciton from the data, predicting that the latter is much larger than that coming from  dissociation into two free polaritons. This suggests that other decay channels such as disorder are important. 

Our theory agrees with the experiments even though it ignores the interaction between excitons with parallel spin. This suggests that even though the mean field shift resulting from this interaction was predicted to be strong~\cite{Rodriguez2016,Delteil2019}, it does not qualitatively affect the Feshbach physics. A precise determination of the polariton density would enable a more systematic comparison between theory and experiment. Additionally, identifying the sources of decay could improve our understanding of the system. In order to obtain even stronger interactions, it would furthermore be desirable to use a resonance with a longer lived Feshbach molecule. Another promising direction is to use polaritons interacting with an electron gas~\cite{Sidler2016,Tan2019}. 

\begin{acknowledgments}
We acknowledge financial support from the Villum Foundation and the Independent Research Fund Denmark - Natural Sciences via Grant No. DFF - 8021-00233B.
\end{acknowledgments}

%%%%%%%%%%%%%%%%%%%%%%%%%%%%%%%%%%%%%%%%%%%%%%%%%%
%%%%%%%%%%%%%%%%%%%%%%%%%%%%%%%%%%%%%%%%%%%%%%%%%%
\appendix
%%%%%%%%%%%%%%%%%%%%%%%%%%%%%%%%%%%%%%%%%%%%%%%%%%
%%%%%%%%%%%%%%%%%%%%%%%%%%%%%%%%%%%%%%%%%%%%%%%%%%

\section{The pair propagator}
\label{app:1}
 Using Eq.~\eqref{eq:Polaritonbasis}
in Eq.~\eqref{eq:Pair} gives
\begin{widetext} 
\begin{gather}
\Pi(\mathbf{k},i\omega_{n})=-T\sum_{\mathbf{q}}\sum_{i\omega'_{n}}\left[\left(\frac{\mathcal{C}^{2}_{\mathbf{k}+\mathbf{q}}}{i\omega_{n}+i\omega'_{\nu}-\varepsilon^{LP}_{\mathbf{k}+\mathbf{q}\downarrow}}+\frac{\mathcal{S}^{2}_{\mathbf{k}+\mathbf{q}}}{i\omega_{nu}+i\omega'_{\nu}-\varepsilon^{UP}_{\mathbf{k}+\mathbf{q}\downarrow}}\right)
%
%\mbox{x}\right. \\ \nonumber\left.
%
\left(\frac{\mathcal{C}^{2}_{-\mathbf{q}}}{-i\omega'_{\nu}-\varepsilon^{LP}_{-\mathbf{q}\uparrow}+\varepsilon^{LP}_{0}}+\frac{\mathcal{S}^{2}_{-\mathbf{q}}}{-i\omega'_{\nu}-\varepsilon^{UP}_{-\mathbf{q}\uparrow}+\varepsilon^{LP}_{0}}\right)\right].
\end{gather}
\end{widetext}
Performing the Matsubara sum yields
\begin{widetext}
\begin{gather}
\Pi(\mathbf{k},i\omega_{n})=\int\frac{{d^{2}\mathbf{q}}}{(2\pi)^{2}}\left\{\mathcal{C}^{2}_{\mathbf{k}+\mathbf{q}}\mathcal{C}^{2}_{-\mathbf{q}}\left[
\frac{1+n_{B}(\varepsilon^{LP}_{\mathbf{k}+\mathbf{q}\downarrow})+n_{B}(\varepsilon^{LP}_{-\mathbf{q}\uparrow}-\varepsilon_{0}^{LP})}{i\omega_{n}-\varepsilon^{LP}_{\mathbf{k}+\mathbf{q}\downarrow}-\varepsilon^{LP}_{-\mathbf{q}\uparrow}+\varepsilon_{0}^{LP}}\right]\right.+
%% II
%% III
\mathcal{C}^{2}_{\mathbf{k}+\mathbf{q}}\mathcal{S}^{2}_{-\mathbf{q}}\left[
\frac{1+n_{B}(\varepsilon^{LP}_{\mathbf{k}+\mathbf{q}\downarrow})+n_{B}(\varepsilon^{UP}_{-\mathbf{q}\uparrow}-\varepsilon_{0}^{LP})}{i\omega_{n}-\varepsilon^{LP}_{\mathbf{k}+\mathbf{q}\downarrow}-\varepsilon^{UP}_{-\mathbf{q}\uparrow}+\varepsilon_{0}^{LP}}\,
\right]  \\ \nonumber
%%IV
+\mathcal{S}^{2}_{\mathbf{k}+\mathbf{q}}\mathcal{C}^{2}_{-\mathbf{q}}\left[
\frac{1+n_{B}(\varepsilon^{UP}_{\mathbf{k}+\mathbf{q}\downarrow})+n_{B}(\varepsilon^{LP}_{-\mathbf{q}\uparrow}-\varepsilon_{0}^{LP})}{i\omega_{n}-\varepsilon^{UP}_{\mathbf{k}+\mathbf{q}\downarrow}-\varepsilon^{LP}_{-\mathbf{q}\uparrow}+\varepsilon_{0}^{LP}}\right]+
%% V
%% VI
\left.\mathcal{S}^{2}_{\mathbf{k}+\mathbf{q}}\mathcal{S}^{2}_{-\mathbf{q}}\left[
\frac{1+n_{B}(\varepsilon^{UP}_{\mathbf{k}+\mathbf{q}\downarrow})+n_{B}(\varepsilon^{UP}_{-\mathbf{q}\uparrow}-\varepsilon_{0}^{LP})}{i\omega_{n}-\varepsilon^{UP}_{\mathbf{k}+\mathbf{q}\downarrow}-\varepsilon^{UP}_{-\mathbf{q}\uparrow}+\varepsilon_{0}^{LP}}
\right]\right\}
\end{gather}
\end{widetext}
with $n_{B}(x)=1/(e^{\beta x}-1)$ being the Bose-Einstein distribution. We consider the zero-temperature case and a vanishing concentration of the impurities so that $n_{B}(x)=0$. Then, the last expression becomes  
\begin{widetext}
\begin{gather}\label{eq:a4}
\Pi(\mathbf{k},i\omega_{n})=\int\frac{{d^{2}\mathbf{q}}}{(2\pi)^{2}}\left\{
\frac{\mathcal{C}^{2}_{\mathbf{k}+\mathbf{q}}\mathcal{C}^{2}_{-\mathbf{q}}}{i\omega_{n}-\varepsilon^{LP}_{\mathbf{k}+\mathbf{q}}-\varepsilon^{LP}_{-\mathbf{q}}+\varepsilon^{LP}_{0}+i\gamma_{0}^{LP}}+
%% III
\frac{\mathcal{C}^{2}_{\mathbf{k}+\mathbf{q}}\mathcal{S}^{2}_{-\mathbf{q}}}{i\omega_{n}-\varepsilon^{LP}_{\mathbf{k}+\mathbf{q}}-\varepsilon^{UP}_{-\mathbf{q}}+\varepsilon^{LP}_{0}+i\gamma_{0}^{UP}}\right.\\ \nonumber 
\left.
%%IV
+\frac{\mathcal{S}^{2}_{\mathbf{k}+\mathbf{q}}\mathcal{C}^{2}_{-\mathbf{q}}}{i\omega_{n}-\varepsilon^{UP}_{\mathbf{k}+\mathbf{q}}-\varepsilon^{LP}_{-\mathbf{q}}+\varepsilon_{0}^{LP}+i\gamma_{0}^{LP}}+
%% VI
\frac{\mathcal{S}^{2}_{\mathbf{k}+\mathbf{q}}\mathcal{S}^{2}_{-\mathbf{q}}}{i\omega_{n}-\varepsilon^{UP}_{\mathbf{k}+\mathbf{q}}-\varepsilon^{UP}_{-\mathbf{q}}+\varepsilon^{LP}_{0}+i\gamma_{0}^{UP}}\right\},
\end{gather} 
\end{widetext}
where we have included a decay for the lower $\gamma_{0}^{LP}$ and upper $\uparrow$-polaritons $\gamma_{0}^{UP}$. In this picture, the pair propagator is expressed as a combination of all the possible polaritonic pairs. It turns out that the Hopfield coefficients tend rapidly to their asymptotic values as $\mathbf{q}$ grows ($\mathcal{C}^{2}_{\mathbf{q}}\rightarrow 1$ and $\mathcal{S}^{2}_{\mathbf{q}}\rightarrow 0$). This means that terms involving $\mathcal{S}_{\mathbf{q}}$ make a small contribution 
to the real part of the pair propagator. They are, however, crucial for calculating the imaginary part since they describe decay of the biexciton into upper polaritons. For zero momentum, the angular integration can be done straightforwardly in Eq.~\eqref{eq:a4} yielding 
\begin{widetext}
\begin{gather}\label{eq:a5}
\Pi(\mathbf{k}=0,i\omega_{n})=\frac{m_{x}}{2\pi}\int_{0}^{E_{\Lambda}}dx\,\left\{
\frac{\mathcal{C}^{4}_{x}}{i\omega_{n}-2\varepsilon^{LP}_{x}-\varepsilon_{0}^{LP}+i\gamma_{0}^{LP}}+
%% III
\frac{\mathcal{C}^{2}_{x}\mathcal{S}^{2}_{x}}{i\omega_{n}-\varepsilon^{LP}_{x}-\varepsilon^{UP}_{x}+\varepsilon_{0}^{LP}+i\gamma_{0}^{UP}}
\right. \nonumber \\ \left.
%%IV
+\frac{\mathcal{S}^{2}_{x}\mathcal{C}^{2}_{x}}{i\omega_{n}-\varepsilon^{UP}_{x}-\varepsilon^{LP}_{x}+\varepsilon_{0}^{LP}+i\gamma_{0}^{LP}}+
%% VI
\frac{\mathcal{S}^{4}_{x}}{i\omega_{n}-2\varepsilon^{UP}_{x}+\varepsilon_{0}^{LP}+i\gamma_{0}^{UP}}\right\}
\end{gather} 
where $E_{\Lambda}=q_{\Lambda}^2/2m_{x}$ is an ultraviolet energy cut-off, which can be taken to infinity after we have renormalized the pair propagator, as explained below. In our calculations we take $\gamma_{0}^{LP}=\gamma_{0}^{UP}$.
\end{widetext}

%%%%%%%%%%%%%%%%%%%%%%%%%%%%%%%%%%%%%%%%%%%%%%%%%%
%%%%%%%%%%%%%%%%%%%%%%%%%%%%%%%%%%%%%%%%%%%%%%%%%%
\section{Scattering matrix and renormalization}
\label{app:2}

The pair propagator and the scattering matrix are related by means of the ladder approximation of the Bethe-Salpeter equation~\cite{Fetter1971}. Expressed in terms of $k\equiv\left(\mathbf{k},i\omega_{n}\right)$, it reads, 
\begin{widetext}
\begin{gather}
\mathcal{T}(k,k';k+q,k-q)=V(k,k',q)
+T\sum_{q'}V(k,k',q')G_{x}^{\downarrow}(k+q')G_{x}^{\uparrow}(k'-q')
\mathcal{T}(k+q,k'-q;k+q-q',k'-q+q').
\end{gather}
\end{widetext}
In our case the interaction $V(k,k',q)$ is static and can be taken to be momentum independent to a good approximation such that $V(\mathbf{k},\mathbf{k}',\mathbf{q})=g$. 
Consequently, we can perform the Matsubara sum and obtain 
\begin{gather} \label{eq:a1}
\mathcal{T}(\mathbf{k},i\omega_{n})=\frac g{1-g\,\Pi(\mathbf{k},i\omega_{n})}.
\end{gather}
We illustrate the scattering matrix calculation in Fig.~\ref{fig:Feynman} (a). As explained in the main text, we  eliminate the UV cut-off by renormalizing our propagator with the binding energy of the biexciton in a vacuum obtained by solving the scattering problem of a $\uparrow\downarrow$ exciton pair. The vacuum pair propagator  is 
\begin{gather}
\Pi_{V}(\mathbf{k},i\omega_{n})=\int\frac{d^{2}\mathbf{q}}{(2\pi)^{2}}\frac{1}{i\omega_{n}-\varepsilon_{\mathbf{k}+\mathbf{q}\downarrow}-\varepsilon_{-\mathbf{q}\uparrow}}.
\end{gather}
For  zero momentum, the integration yields
\begin{gather} 
\Pi_{V}(i\omega_{n})=-\frac{m_{x}}{4\pi}\left[\log\left(\frac{2E_{\Lambda}-i\omega_{n}}{-i\omega_{n}}\right)+i\pi\right],
\end{gather}
where $E_{\Lambda}$ is the ultraviolet cut-off energy. 
We identify the binding energy of the biexciton as the real pole of the zero-momentum scattering matrix $g^{-1}=\mbox{Re}\Pi^{V}(E_{B})$. By substituting this into Eq.~\eqref{eq:a1} we arrive at the re-normalized scattering matrix
\begin{gather}
\mathcal{T}(\mathbf{k},i\omega_{n})=\frac{1}{\mbox{Re}\Pi_{V}(E_{B})-\Pi(\mathbf{k},i\omega_{n})}.
\end{gather}
Finally, the self energy of the impurity is 
$\Sigma(\mathbf{k},\omega)=n_{x\uparrow}\mathcal{T}(\mathbf{k},i\omega_{n})$, where 
$n_{x\uparrow}$ is the density of $\uparrow$-excitons in the BEC. 
%%%%%%%%%%%%%%%%%%%%%%%%%%%%%%%%%%%%%%%%%%%%%%%%%%
%%%%%%%%%%%%%%%%%%%%%%%%%%%%%%%%%%%%%%%%%%%%%%%%%%

%%%%%%%%%%%%%%%%%%%%%%%%%%%%%%%%%%%%%%%%%%%%%%%%%%
\bibliographystyle{apsrev4-1}
\bibliography{EP_Notes}

\begin{thebibliography}{52}%
\makeatletter
\providecommand \@ifxundefined [1]{%
 \@ifx{#1\undefined}
}%
\providecommand \@ifnum [1]{%
 \ifnum #1\expandafter \@firstoftwo
 \else \expandafter \@secondoftwo
 \fi
}%
\providecommand \@ifx [1]{%
 \ifx #1\expandafter \@firstoftwo
 \else \expandafter \@secondoftwo
 \fi
}%
\providecommand \natexlab [1]{#1}%
\providecommand \enquote  [1]{``#1''}%
\providecommand \bibnamefont  [1]{#1}%
\providecommand \bibfnamefont [1]{#1}%
\providecommand \citenamefont [1]{#1}%
\providecommand \href@noop [0]{\@secondoftwo}%
\providecommand \href [0]{\begingroup \@sanitize@url \@href}%
\providecommand \@href[1]{\@@startlink{#1}\@@href}%
\providecommand \@@href[1]{\endgroup#1\@@endlink}%
\providecommand \@sanitize@url [0]{\catcode `\\12\catcode `\$12\catcode
  `\&12\catcode `\#12\catcode `\^12\catcode `\_12\catcode `\%12\relax}%
\providecommand \@@startlink[1]{}%
\providecommand \@@endlink[0]{}%
\providecommand \url  [0]{\begingroup\@sanitize@url \@url }%
\providecommand \@url [1]{\endgroup\@href {#1}{\urlprefix }}%
\providecommand \urlprefix  [0]{URL }%
\providecommand \Eprint [0]{\href }%
\providecommand \doibase [0]{http://dx.doi.org/}%
\providecommand \selectlanguage [0]{\@gobble}%
\providecommand \bibinfo  [0]{\@secondoftwo}%
\providecommand \bibfield  [0]{\@secondoftwo}%
\providecommand \translation [1]{[#1]}%
\providecommand \BibitemOpen [0]{}%
\providecommand \bibitemStop [0]{}%
\providecommand \bibitemNoStop [0]{.\EOS\space}%
\providecommand \EOS [0]{\spacefactor3000\relax}%
\providecommand \BibitemShut  [1]{\csname bibitem#1\endcsname}%
\let\auto@bib@innerbib\@empty
%</preamble>
\bibitem [{\citenamefont {Hopfield}(1958)}]{Hopfield1958}%
  \BibitemOpen
  \bibfield  {author} {\bibinfo {author} {\bibfnamefont {J.~J.}\ \bibnamefont
  {Hopfield}},\ }\href {\doibase 10.1103/PhysRev.112.1555} {\bibfield
  {journal} {\bibinfo  {journal} {Phys. Rev.}\ }\textbf {\bibinfo {volume}
  {112}},\ \bibinfo {pages} {1555} (\bibinfo {year} {1958})}\BibitemShut
  {NoStop}%
\bibitem [{\citenamefont {Weisbuch}\ \emph {et~al.}(1992)\citenamefont
  {Weisbuch}, \citenamefont {Nishioka}, \citenamefont {Ishikawa},\ and\
  \citenamefont {Arakawa}}]{Weisbuch1992}%
  \BibitemOpen
  \bibfield  {author} {\bibinfo {author} {\bibfnamefont {C.}~\bibnamefont
  {Weisbuch}}, \bibinfo {author} {\bibfnamefont {M.}~\bibnamefont {Nishioka}},
  \bibinfo {author} {\bibfnamefont {A.}~\bibnamefont {Ishikawa}}, \ and\
  \bibinfo {author} {\bibfnamefont {Y.}~\bibnamefont {Arakawa}},\ }\href
  {\doibase 10.1103/PhysRevLett.69.3314} {\bibfield  {journal} {\bibinfo
  {journal} {Phys. Rev. Lett.}\ }\textbf {\bibinfo {volume} {69}},\ \bibinfo
  {pages} {3314} (\bibinfo {year} {1992})}\BibitemShut {NoStop}%
\bibitem [{\citenamefont {Carusotto}\ and\ \citenamefont
  {Ciuti}(2013)}]{Carusotto2013}%
  \BibitemOpen
  \bibfield  {author} {\bibinfo {author} {\bibfnamefont {I.}~\bibnamefont
  {Carusotto}}\ and\ \bibinfo {author} {\bibfnamefont {C.}~\bibnamefont
  {Ciuti}},\ }\href {\doibase 10.1103/RevModPhys.85.299} {\bibfield  {journal}
  {\bibinfo  {journal} {Rev. Mod. Phys.}\ }\textbf {\bibinfo {volume} {85}},\
  \bibinfo {pages} {299} (\bibinfo {year} {2013})}\BibitemShut {NoStop}%
\bibitem [{\citenamefont {Kavokin}\ \emph {et~al.}(2017)\citenamefont
  {Kavokin}, \citenamefont {Baumberg}, \citenamefont {Malpuech},\ and\
  \citenamefont {Laussy}}]{Kavokin2017}%
  \BibitemOpen
  \bibfield  {author} {\bibinfo {author} {\bibfnamefont {A.~V.}\ \bibnamefont
  {Kavokin}}, \bibinfo {author} {\bibfnamefont {J.~J.}\ \bibnamefont
  {Baumberg}}, \bibinfo {author} {\bibfnamefont {G.}~\bibnamefont {Malpuech}},
  \ and\ \bibinfo {author} {\bibfnamefont {F.~P.}\ \bibnamefont {Laussy}},\
  }\href@noop {} {\emph {\bibinfo {title} {Microcavities}}},\ Series on
  Semiconductor Science and Technology\ (\bibinfo  {publisher} {Oxford
  University Press},\ \bibinfo {year} {2017})\BibitemShut {NoStop}%
\bibitem [{\citenamefont {Kasprzak}\ \emph {et~al.}(2006)\citenamefont
  {Kasprzak}, \citenamefont {Richard}, \citenamefont {Kundermann},
  \citenamefont {Baas}, \citenamefont {Jeambrun}, \citenamefont {Keeling},
  \citenamefont {Marchetti}, \citenamefont {Szyma\'nska}, \citenamefont
  {Andr\'e}, \citenamefont {Staehli}, \citenamefont {Savona}, \citenamefont
  {Littlewood}, \citenamefont {Deveaud},\ and\ \citenamefont
  {Dang}}]{Kasprzak2006}%
  \BibitemOpen
  \bibfield  {author} {\bibinfo {author} {\bibfnamefont {J.}~\bibnamefont
  {Kasprzak}}, \bibinfo {author} {\bibfnamefont {M.}~\bibnamefont {Richard}},
  \bibinfo {author} {\bibfnamefont {S.}~\bibnamefont {Kundermann}}, \bibinfo
  {author} {\bibfnamefont {A.}~\bibnamefont {Baas}}, \bibinfo {author}
  {\bibfnamefont {P.}~\bibnamefont {Jeambrun}}, \bibinfo {author}
  {\bibfnamefont {J.~M.~J.}\ \bibnamefont {Keeling}}, \bibinfo {author}
  {\bibfnamefont {F.~M.}\ \bibnamefont {Marchetti}}, \bibinfo {author}
  {\bibfnamefont {M.~H.}\ \bibnamefont {Szyma\'nska}}, \bibinfo {author}
  {\bibfnamefont {R.}~\bibnamefont {Andr\'e}}, \bibinfo {author} {\bibfnamefont
  {J.~L.}\ \bibnamefont {Staehli}}, \bibinfo {author} {\bibfnamefont
  {V.}~\bibnamefont {Savona}}, \bibinfo {author} {\bibfnamefont {P.~B.}\
  \bibnamefont {Littlewood}}, \bibinfo {author} {\bibfnamefont
  {B.}~\bibnamefont {Deveaud}}, \ and\ \bibinfo {author} {\bibfnamefont
  {L.~S.}\ \bibnamefont {Dang}},\ }\href {\doibase 10.1038/nature05131}
  {\bibfield  {journal} {\bibinfo  {journal} {Nature}\ }\textbf {\bibinfo
  {volume} {443}},\ \bibinfo {pages} {409} (\bibinfo {year}
  {2006})}\BibitemShut {NoStop}%
\bibitem [{\citenamefont {Wouters}\ and\ \citenamefont
  {Carusotto}(2007)}]{Wouters2007a}%
  \BibitemOpen
  \bibfield  {author} {\bibinfo {author} {\bibfnamefont {M.}~\bibnamefont
  {Wouters}}\ and\ \bibinfo {author} {\bibfnamefont {I.}~\bibnamefont
  {Carusotto}},\ }\href {\doibase 10.1103/PhysRevLett.99.140402} {\bibfield
  {journal} {\bibinfo  {journal} {Phys. Rev. Lett.}\ }\textbf {\bibinfo
  {volume} {99}},\ \bibinfo {pages} {140402} (\bibinfo {year}
  {2007})}\BibitemShut {NoStop}%
\bibitem [{\citenamefont {Amo}\ \emph {et~al.}(2009)\citenamefont {Amo},
  \citenamefont {Lefr\`ere}, \citenamefont {Pigeon}, \citenamefont {Adrados},
  \citenamefont {Ciuti}, \citenamefont {Carusotto}, \citenamefont {Houdr\'e},
  \citenamefont {Giacobino},\ and\ \citenamefont {Bramati}}]{Amo2009}%
  \BibitemOpen
  \bibfield  {author} {\bibinfo {author} {\bibfnamefont {A.}~\bibnamefont
  {Amo}}, \bibinfo {author} {\bibfnamefont {J.}~\bibnamefont {Lefr\`ere}},
  \bibinfo {author} {\bibfnamefont {S.}~\bibnamefont {Pigeon}}, \bibinfo
  {author} {\bibfnamefont {C.}~\bibnamefont {Adrados}}, \bibinfo {author}
  {\bibfnamefont {C.}~\bibnamefont {Ciuti}}, \bibinfo {author} {\bibfnamefont
  {I.}~\bibnamefont {Carusotto}}, \bibinfo {author} {\bibfnamefont
  {R.}~\bibnamefont {Houdr\'e}}, \bibinfo {author} {\bibfnamefont
  {E.}~\bibnamefont {Giacobino}}, \ and\ \bibinfo {author} {\bibfnamefont
  {A.}~\bibnamefont {Bramati}},\ }\href {https://doi.org/10.1038/nphys1364}
  {\bibfield  {journal} {\bibinfo  {journal} {Nature Physics}\ }\textbf
  {\bibinfo {volume} {5}},\ \bibinfo {pages} {805 EP } (\bibinfo {year}
  {2009})}\BibitemShut {NoStop}%
\bibitem [{\citenamefont {Deng}\ \emph {et~al.}(2010)\citenamefont {Deng},
  \citenamefont {Haug},\ and\ \citenamefont {Yamamoto}}]{Deng2010}%
  \BibitemOpen
  \bibfield  {author} {\bibinfo {author} {\bibfnamefont {H.}~\bibnamefont
  {Deng}}, \bibinfo {author} {\bibfnamefont {H.}~\bibnamefont {Haug}}, \ and\
  \bibinfo {author} {\bibfnamefont {Y.}~\bibnamefont {Yamamoto}},\ }\href
  {\doibase 10.1103/RevModPhys.82.1489} {\bibfield  {journal} {\bibinfo
  {journal} {Rev. Mod. Phys.}\ }\textbf {\bibinfo {volume} {82}},\ \bibinfo
  {pages} {1489} (\bibinfo {year} {2010})}\BibitemShut {NoStop}%
\bibitem [{\citenamefont {Kohnle}\ \emph {et~al.}(2011)\citenamefont {Kohnle},
  \citenamefont {L\'eger}, \citenamefont {Wouters}, \citenamefont {Richard},
  \citenamefont {Portella-Oberli},\ and\ \citenamefont
  {Deveaud-Pl\'edran}}]{Kohnle2011}%
  \BibitemOpen
  \bibfield  {author} {\bibinfo {author} {\bibfnamefont {V.}~\bibnamefont
  {Kohnle}}, \bibinfo {author} {\bibfnamefont {Y.}~\bibnamefont {L\'eger}},
  \bibinfo {author} {\bibfnamefont {M.}~\bibnamefont {Wouters}}, \bibinfo
  {author} {\bibfnamefont {M.}~\bibnamefont {Richard}}, \bibinfo {author}
  {\bibfnamefont {M.~T.}\ \bibnamefont {Portella-Oberli}}, \ and\ \bibinfo
  {author} {\bibfnamefont {B.}~\bibnamefont {Deveaud-Pl\'edran}},\ }\href
  {\doibase 10.1103/PhysRevLett.106.255302} {\bibfield  {journal} {\bibinfo
  {journal} {Phys. Rev. Lett.}\ }\textbf {\bibinfo {volume} {106}},\ \bibinfo
  {pages} {255302} (\bibinfo {year} {2011})}\BibitemShut {NoStop}%
\bibitem [{\citenamefont {Kohnle}\ \emph {et~al.}(2012)\citenamefont {Kohnle},
  \citenamefont {L\'eger}, \citenamefont {Wouters}, \citenamefont {Richard},
  \citenamefont {Portella-Oberli},\ and\ \citenamefont {Deveaud}}]{Kohnle2012}%
  \BibitemOpen
  \bibfield  {author} {\bibinfo {author} {\bibfnamefont {V.}~\bibnamefont
  {Kohnle}}, \bibinfo {author} {\bibfnamefont {Y.}~\bibnamefont {L\'eger}},
  \bibinfo {author} {\bibfnamefont {M.}~\bibnamefont {Wouters}}, \bibinfo
  {author} {\bibfnamefont {M.}~\bibnamefont {Richard}}, \bibinfo {author}
  {\bibfnamefont {M.~T.}\ \bibnamefont {Portella-Oberli}}, \ and\ \bibinfo
  {author} {\bibfnamefont {B.}~\bibnamefont {Deveaud}},\ }\href {\doibase
  10.1103/PhysRevB.86.064508} {\bibfield  {journal} {\bibinfo  {journal} {Phys.
  Rev. B}\ }\textbf {\bibinfo {volume} {86}},\ \bibinfo {pages} {064508}
  (\bibinfo {year} {2012})}\BibitemShut {NoStop}%
\bibitem [{\citenamefont {Lagoudakis}\ \emph {et~al.}(2008)\citenamefont
  {Lagoudakis}, \citenamefont {Wouters}, \citenamefont {Richard}, \citenamefont
  {Baas}, \citenamefont {Carusotto}, \citenamefont {Andr{\'e}}, \citenamefont
  {Dang},\ and\ \citenamefont {Deveaud-Pl{\'e}dran}}]{Lagoudakis2008}%
  \BibitemOpen
  \bibfield  {author} {\bibinfo {author} {\bibfnamefont {K.~G.}\ \bibnamefont
  {Lagoudakis}}, \bibinfo {author} {\bibfnamefont {M.}~\bibnamefont {Wouters}},
  \bibinfo {author} {\bibfnamefont {M.}~\bibnamefont {Richard}}, \bibinfo
  {author} {\bibfnamefont {A.}~\bibnamefont {Baas}}, \bibinfo {author}
  {\bibfnamefont {I.}~\bibnamefont {Carusotto}}, \bibinfo {author}
  {\bibfnamefont {R.}~\bibnamefont {Andr{\'e}}}, \bibinfo {author}
  {\bibfnamefont {L.~S.}\ \bibnamefont {Dang}}, \ and\ \bibinfo {author}
  {\bibfnamefont {B.}~\bibnamefont {Deveaud-Pl{\'e}dran}},\ }\href
  {https://doi.org/10.1038/nphys1051} {\bibfield  {journal} {\bibinfo
  {journal} {Nature Physics}\ }\textbf {\bibinfo {volume} {4}},\ \bibinfo
  {pages} {706 EP } (\bibinfo {year} {2008})}\BibitemShut {NoStop}%
\bibitem [{\citenamefont {St-Jean}\ \emph {et~al.}(2017)\citenamefont
  {St-Jean}, \citenamefont {Goblot}, \citenamefont {Galopin}, \citenamefont
  {Lema{\^\i}tre}, \citenamefont {Ozawa}, \citenamefont {Le~Gratiet},
  \citenamefont {Sagnes}, \citenamefont {Bloch},\ and\ \citenamefont
  {Amo}}]{St-Jean2017}%
  \BibitemOpen
  \bibfield  {author} {\bibinfo {author} {\bibfnamefont {P.}~\bibnamefont
  {St-Jean}}, \bibinfo {author} {\bibfnamefont {V.}~\bibnamefont {Goblot}},
  \bibinfo {author} {\bibfnamefont {E.}~\bibnamefont {Galopin}}, \bibinfo
  {author} {\bibfnamefont {A.}~\bibnamefont {Lema{\^\i}tre}}, \bibinfo {author}
  {\bibfnamefont {T.}~\bibnamefont {Ozawa}}, \bibinfo {author} {\bibfnamefont
  {L.}~\bibnamefont {Le~Gratiet}}, \bibinfo {author} {\bibfnamefont
  {I.}~\bibnamefont {Sagnes}}, \bibinfo {author} {\bibfnamefont
  {J.}~\bibnamefont {Bloch}}, \ and\ \bibinfo {author} {\bibfnamefont
  {A.}~\bibnamefont {Amo}},\ }\href {\doibase 10.1038/s41566-017-0006-2}
  {\bibfield  {journal} {\bibinfo  {journal} {Nature Photonics}\ }\textbf
  {\bibinfo {volume} {11}},\ \bibinfo {pages} {651} (\bibinfo {year}
  {2017})}\BibitemShut {NoStop}%
\bibitem [{\citenamefont {Klembt}\ \emph {et~al.}(2018)\citenamefont {Klembt},
  \citenamefont {Harder}, \citenamefont {Egorov}, \citenamefont {Winkler},
  \citenamefont {Ge}, \citenamefont {Bandres}, \citenamefont {Emmerling},
  \citenamefont {Worschech}, \citenamefont {Liew}, \citenamefont {Segev},
  \citenamefont {Schneider},\ and\ \citenamefont {H{\"o}fling}}]{Klembt2018}%
  \BibitemOpen
  \bibfield  {author} {\bibinfo {author} {\bibfnamefont {S.}~\bibnamefont
  {Klembt}}, \bibinfo {author} {\bibfnamefont {T.~H.}\ \bibnamefont {Harder}},
  \bibinfo {author} {\bibfnamefont {O.~A.}\ \bibnamefont {Egorov}}, \bibinfo
  {author} {\bibfnamefont {K.}~\bibnamefont {Winkler}}, \bibinfo {author}
  {\bibfnamefont {R.}~\bibnamefont {Ge}}, \bibinfo {author} {\bibfnamefont
  {M.~A.}\ \bibnamefont {Bandres}}, \bibinfo {author} {\bibfnamefont
  {M.}~\bibnamefont {Emmerling}}, \bibinfo {author} {\bibfnamefont
  {L.}~\bibnamefont {Worschech}}, \bibinfo {author} {\bibfnamefont {T.~C.~H.}\
  \bibnamefont {Liew}}, \bibinfo {author} {\bibfnamefont {M.}~\bibnamefont
  {Segev}}, \bibinfo {author} {\bibfnamefont {C.}~\bibnamefont {Schneider}}, \
  and\ \bibinfo {author} {\bibfnamefont {S.}~\bibnamefont {H{\"o}fling}},\
  }\href {\doibase 10.1038/s41586-018-0601-5} {\bibfield  {journal} {\bibinfo
  {journal} {Nature}\ }\textbf {\bibinfo {volume} {562}},\ \bibinfo {pages}
  {552} (\bibinfo {year} {2018})}\BibitemShut {NoStop}%
\bibitem [{\citenamefont {Sanvitto}\ and\ \citenamefont
  {K{\'e}na-Cohen}(2016)}]{Sanvitto2016}%
  \BibitemOpen
  \bibfield  {author} {\bibinfo {author} {\bibfnamefont {D.}~\bibnamefont
  {Sanvitto}}\ and\ \bibinfo {author} {\bibfnamefont {S.}~\bibnamefont
  {K{\'e}na-Cohen}},\ }\href {https://doi.org/10.1038/nmat4668} {\bibfield
  {journal} {\bibinfo  {journal} {Nature Materials}\ }\textbf {\bibinfo
  {volume} {15}},\ \bibinfo {pages} {1061 EP } (\bibinfo {year}
  {2016})}\BibitemShut {NoStop}%
\bibitem [{\citenamefont {Mu{\~n}oz-Matutano}\ \emph
  {et~al.}(2019)\citenamefont {Mu{\~n}oz-Matutano}, \citenamefont {Wood},
  \citenamefont {Johnsson}, \citenamefont {Vidal}, \citenamefont {Baragiola},
  \citenamefont {Reinhard}, \citenamefont {Lema{\^\i}tre}, \citenamefont
  {Bloch}, \citenamefont {Amo}, \citenamefont {Nogues}, \citenamefont {Besga},
  \citenamefont {Richard},\ and\ \citenamefont {Volz}}]{Munoz-Matutano2019}%
  \BibitemOpen
  \bibfield  {author} {\bibinfo {author} {\bibfnamefont {G.}~\bibnamefont
  {Mu{\~n}oz-Matutano}}, \bibinfo {author} {\bibfnamefont {A.}~\bibnamefont
  {Wood}}, \bibinfo {author} {\bibfnamefont {M.}~\bibnamefont {Johnsson}},
  \bibinfo {author} {\bibfnamefont {X.}~\bibnamefont {Vidal}}, \bibinfo
  {author} {\bibfnamefont {B.~Q.}\ \bibnamefont {Baragiola}}, \bibinfo {author}
  {\bibfnamefont {A.}~\bibnamefont {Reinhard}}, \bibinfo {author}
  {\bibfnamefont {A.}~\bibnamefont {Lema{\^\i}tre}}, \bibinfo {author}
  {\bibfnamefont {J.}~\bibnamefont {Bloch}}, \bibinfo {author} {\bibfnamefont
  {A.}~\bibnamefont {Amo}}, \bibinfo {author} {\bibfnamefont {G.}~\bibnamefont
  {Nogues}}, \bibinfo {author} {\bibfnamefont {B.}~\bibnamefont {Besga}},
  \bibinfo {author} {\bibfnamefont {M.}~\bibnamefont {Richard}}, \ and\
  \bibinfo {author} {\bibfnamefont {T.}~\bibnamefont {Volz}},\ }\href {\doibase
  10.1038/s41563-019-0281-z} {\bibfield  {journal} {\bibinfo  {journal} {Nature
  Materials}\ }\textbf {\bibinfo {volume} {18}},\ \bibinfo {pages} {213}
  (\bibinfo {year} {2019})}\BibitemShut {NoStop}%
\bibitem [{\citenamefont {Delteil}\ \emph {et~al.}(2019)\citenamefont
  {Delteil}, \citenamefont {Fink}, \citenamefont {Schade}, \citenamefont
  {H{\"o}fling}, \citenamefont {Schneider},\ and\ \citenamefont {{\.I}mamo{\u
  g}lu}}]{Delteil2019}%
  \BibitemOpen
  \bibfield  {author} {\bibinfo {author} {\bibfnamefont {A.}~\bibnamefont
  {Delteil}}, \bibinfo {author} {\bibfnamefont {T.}~\bibnamefont {Fink}},
  \bibinfo {author} {\bibfnamefont {A.}~\bibnamefont {Schade}}, \bibinfo
  {author} {\bibfnamefont {S.}~\bibnamefont {H{\"o}fling}}, \bibinfo {author}
  {\bibfnamefont {C.}~\bibnamefont {Schneider}}, \ and\ \bibinfo {author}
  {\bibfnamefont {A.}~\bibnamefont {{\.I}mamo{\u g}lu}},\ }\href {\doibase
  10.1038/s41563-019-0282-y} {\bibfield  {journal} {\bibinfo  {journal} {Nature
  Materials}\ }\textbf {\bibinfo {volume} {18}},\ \bibinfo {pages} {219}
  (\bibinfo {year} {2019})}\BibitemShut {NoStop}%
\bibitem [{\citenamefont {Kn{\"u}Eppel}\ \emph {et~al.}(2019)\citenamefont
  {Kn{\"u}Eppel}, \citenamefont {Ravets}, \citenamefont {Kroner}, \citenamefont
  {F{\"a}lt}, \citenamefont {Wegscheider},\ and\ \citenamefont
  {Imamoglu}}]{Knuppel2019}%
  \BibitemOpen
  \bibfield  {author} {\bibinfo {author} {\bibfnamefont {P.}~\bibnamefont
  {Kn{\"u}Eppel}}, \bibinfo {author} {\bibfnamefont {S.}~\bibnamefont
  {Ravets}}, \bibinfo {author} {\bibfnamefont {M.}~\bibnamefont {Kroner}},
  \bibinfo {author} {\bibfnamefont {S.}~\bibnamefont {F{\"a}lt}}, \bibinfo
  {author} {\bibfnamefont {W.}~\bibnamefont {Wegscheider}}, \ and\ \bibinfo
  {author} {\bibfnamefont {A.}~\bibnamefont {Imamoglu}},\ }\href {\doibase
  10.1038/s41586-019-1356-3} {\bibfield  {journal} {\bibinfo  {journal}
  {Nature}\ }\textbf {\bibinfo {volume} {572}},\ \bibinfo {pages} {91}
  (\bibinfo {year} {2019})}\BibitemShut {NoStop}%
\bibitem [{\citenamefont {Cristofolini}\ \emph {et~al.}(2012)\citenamefont
  {Cristofolini}, \citenamefont {Christmann}, \citenamefont {Tsintzos},
  \citenamefont {Deligeorgis}, \citenamefont {Konstantinidis}, \citenamefont
  {Hatzopoulos}, \citenamefont {Savvidis},\ and\ \citenamefont
  {Baumberg}}]{Cristofolini2012}%
  \BibitemOpen
  \bibfield  {author} {\bibinfo {author} {\bibfnamefont {P.}~\bibnamefont
  {Cristofolini}}, \bibinfo {author} {\bibfnamefont {G.}~\bibnamefont
  {Christmann}}, \bibinfo {author} {\bibfnamefont {S.~I.}\ \bibnamefont
  {Tsintzos}}, \bibinfo {author} {\bibfnamefont {G.}~\bibnamefont
  {Deligeorgis}}, \bibinfo {author} {\bibfnamefont {G.}~\bibnamefont
  {Konstantinidis}}, \bibinfo {author} {\bibfnamefont {Z.}~\bibnamefont
  {Hatzopoulos}}, \bibinfo {author} {\bibfnamefont {P.~G.}\ \bibnamefont
  {Savvidis}}, \ and\ \bibinfo {author} {\bibfnamefont {J.~J.}\ \bibnamefont
  {Baumberg}},\ }\href {\doibase 10.1126/science.1219010} {\bibfield  {journal}
  {\bibinfo  {journal} {Science}\ }\textbf {\bibinfo {volume} {336}},\ \bibinfo
  {pages} {704} (\bibinfo {year} {2012})}\BibitemShut {NoStop}%
\bibitem [{\citenamefont {Byrnes}\ \emph {et~al.}(2014)\citenamefont {Byrnes},
  \citenamefont {Kolmakov}, \citenamefont {Kezerashvili},\ and\ \citenamefont
  {Yamamoto}}]{Byrnes2014}%
  \BibitemOpen
  \bibfield  {author} {\bibinfo {author} {\bibfnamefont {T.}~\bibnamefont
  {Byrnes}}, \bibinfo {author} {\bibfnamefont {G.~V.}\ \bibnamefont
  {Kolmakov}}, \bibinfo {author} {\bibfnamefont {R.~Y.}\ \bibnamefont
  {Kezerashvili}}, \ and\ \bibinfo {author} {\bibfnamefont {Y.}~\bibnamefont
  {Yamamoto}},\ }\href {\doibase 10.1103/PhysRevB.90.125314} {\bibfield
  {journal} {\bibinfo  {journal} {Phys. Rev. B}\ }\textbf {\bibinfo {volume}
  {90}},\ \bibinfo {pages} {125314} (\bibinfo {year} {2014})}\BibitemShut
  {NoStop}%
\bibitem [{\citenamefont {Rosenberg}\ \emph {et~al.}(2018)\citenamefont
  {Rosenberg}, \citenamefont {Liran}, \citenamefont {Mazuz-Harpaz},
  \citenamefont {West}, \citenamefont {Pfeiffer},\ and\ \citenamefont
  {Rapaport}}]{Rosenberg2018}%
  \BibitemOpen
  \bibfield  {author} {\bibinfo {author} {\bibfnamefont {I.}~\bibnamefont
  {Rosenberg}}, \bibinfo {author} {\bibfnamefont {D.}~\bibnamefont {Liran}},
  \bibinfo {author} {\bibfnamefont {Y.}~\bibnamefont {Mazuz-Harpaz}}, \bibinfo
  {author} {\bibfnamefont {K.}~\bibnamefont {West}}, \bibinfo {author}
  {\bibfnamefont {L.}~\bibnamefont {Pfeiffer}}, \ and\ \bibinfo {author}
  {\bibfnamefont {R.}~\bibnamefont {Rapaport}},\ }\href {\doibase
  10.1126/sciadv.aat8880} {\bibfield  {journal} {\bibinfo  {journal} {Science
  Advances}\ }\textbf {\bibinfo {volume} {4}},\ \bibinfo {pages} {eaat8880}
  (\bibinfo {year} {2018})}\BibitemShut {NoStop}%
\bibitem [{\citenamefont {Togan}\ \emph {et~al.}(2018)\citenamefont {Togan},
  \citenamefont {Lim}, \citenamefont {Faelt}, \citenamefont {Wegscheider},\
  and\ \citenamefont {Imamoglu}}]{Togan2018}%
  \BibitemOpen
  \bibfield  {author} {\bibinfo {author} {\bibfnamefont {E.}~\bibnamefont
  {Togan}}, \bibinfo {author} {\bibfnamefont {H.-T.}\ \bibnamefont {Lim}},
  \bibinfo {author} {\bibfnamefont {S.}~\bibnamefont {Faelt}}, \bibinfo
  {author} {\bibfnamefont {W.}~\bibnamefont {Wegscheider}}, \ and\ \bibinfo
  {author} {\bibfnamefont {A.}~\bibnamefont {Imamoglu}},\ }\href {\doibase
  10.1103/PhysRevLett.121.227402} {\bibfield  {journal} {\bibinfo  {journal}
  {Phys. Rev. Lett.}\ }\textbf {\bibinfo {volume} {121}},\ \bibinfo {pages}
  {227402} (\bibinfo {year} {2018})}\BibitemShut {NoStop}%
\bibitem [{\citenamefont {Takemura}\ \emph
  {et~al.}(2014{\natexlab{a}})\citenamefont {Takemura}, \citenamefont
  {Trebaol}, \citenamefont {Wouters}, \citenamefont {Portella-Oberli},\ and\
  \citenamefont {Deveaud}}]{Takemura2014}%
  \BibitemOpen
  \bibfield  {author} {\bibinfo {author} {\bibfnamefont {N.}~\bibnamefont
  {Takemura}}, \bibinfo {author} {\bibfnamefont {S.}~\bibnamefont {Trebaol}},
  \bibinfo {author} {\bibfnamefont {M.}~\bibnamefont {Wouters}}, \bibinfo
  {author} {\bibfnamefont {M.~T.}\ \bibnamefont {Portella-Oberli}}, \ and\
  \bibinfo {author} {\bibfnamefont {B.}~\bibnamefont {Deveaud}},\ }\href
  {https://doi.org/10.1038/nphys2999} {\bibfield  {journal} {\bibinfo
  {journal} {Nature Physics}\ }\textbf {\bibinfo {volume} {10}},\ \bibinfo
  {pages} {500 EP} (\bibinfo {year} {2014}{\natexlab{a}})}\BibitemShut
  {NoStop}%
\bibitem [{\citenamefont {Takemura}\ \emph
  {et~al.}(2014{\natexlab{b}})\citenamefont {Takemura}, \citenamefont
  {Trebaol}, \citenamefont {Wouters}, \citenamefont {Portella-Oberli},\ and\
  \citenamefont {Deveaud}}]{Takemura2014b}%
  \BibitemOpen
  \bibfield  {author} {\bibinfo {author} {\bibfnamefont {N.}~\bibnamefont
  {Takemura}}, \bibinfo {author} {\bibfnamefont {S.}~\bibnamefont {Trebaol}},
  \bibinfo {author} {\bibfnamefont {M.}~\bibnamefont {Wouters}}, \bibinfo
  {author} {\bibfnamefont {M.~T.}\ \bibnamefont {Portella-Oberli}}, \ and\
  \bibinfo {author} {\bibfnamefont {B.}~\bibnamefont {Deveaud}},\ }\href
  {\doibase 10.1103/PhysRevB.90.195307} {\bibfield  {journal} {\bibinfo
  {journal} {Phys. Rev. B}\ }\textbf {\bibinfo {volume} {90}},\ \bibinfo
  {pages} {195307} (\bibinfo {year} {2014}{\natexlab{b}})}\BibitemShut
  {NoStop}%
\bibitem [{\citenamefont {Takemura}\ \emph {et~al.}(2017)\citenamefont
  {Takemura}, \citenamefont {Anderson}, \citenamefont {Navadeh-Toupchi},
  \citenamefont {Oberli}, \citenamefont {Portella-Oberli},\ and\ \citenamefont
  {Deveaud}}]{Takemura2017}%
  \BibitemOpen
  \bibfield  {author} {\bibinfo {author} {\bibfnamefont {N.}~\bibnamefont
  {Takemura}}, \bibinfo {author} {\bibfnamefont {M.~D.}\ \bibnamefont
  {Anderson}}, \bibinfo {author} {\bibfnamefont {M.}~\bibnamefont
  {Navadeh-Toupchi}}, \bibinfo {author} {\bibfnamefont {D.~Y.}\ \bibnamefont
  {Oberli}}, \bibinfo {author} {\bibfnamefont {M.~T.}\ \bibnamefont
  {Portella-Oberli}}, \ and\ \bibinfo {author} {\bibfnamefont {B.}~\bibnamefont
  {Deveaud}},\ }\href {\doibase 10.1103/PhysRevB.95.205303} {\bibfield
  {journal} {\bibinfo  {journal} {Phys. Rev. B}\ }\textbf {\bibinfo {volume}
  {95}},\ \bibinfo {pages} {205303} (\bibinfo {year} {2017})}\BibitemShut
  {NoStop}%
\bibitem [{\citenamefont {Navadeh-Toupchi}\ \emph {et~al.}(2019)\citenamefont
  {Navadeh-Toupchi}, \citenamefont {Takemura}, \citenamefont {Anderson},
  \citenamefont {Oberli},\ and\ \citenamefont {Portella-Oberli}}]{Navadeh2019}%
  \BibitemOpen
  \bibfield  {author} {\bibinfo {author} {\bibfnamefont {M.}~\bibnamefont
  {Navadeh-Toupchi}}, \bibinfo {author} {\bibfnamefont {N.}~\bibnamefont
  {Takemura}}, \bibinfo {author} {\bibfnamefont {M.~D.}\ \bibnamefont
  {Anderson}}, \bibinfo {author} {\bibfnamefont {D.~Y.}\ \bibnamefont
  {Oberli}}, \ and\ \bibinfo {author} {\bibfnamefont {M.~T.}\ \bibnamefont
  {Portella-Oberli}},\ }\href {\doibase 10.1103/PhysRevLett.122.047402}
  {\bibfield  {journal} {\bibinfo  {journal} {Phys. Rev. Lett.}\ }\textbf
  {\bibinfo {volume} {122}},\ \bibinfo {pages} {047402} (\bibinfo {year}
  {2019})}\BibitemShut {NoStop}%
\bibitem [{\citenamefont {Rocca}\ \emph {et~al.}(1998)\citenamefont {Rocca},
  \citenamefont {Bassani},\ and\ \citenamefont {Agranovich}}]{LaRocca1998}%
  \BibitemOpen
  \bibfield  {author} {\bibinfo {author} {\bibfnamefont {G.~C.~L.}\
  \bibnamefont {Rocca}}, \bibinfo {author} {\bibfnamefont {F.}~\bibnamefont
  {Bassani}}, \ and\ \bibinfo {author} {\bibfnamefont {V.~M.}\ \bibnamefont
  {Agranovich}},\ }\href {\doibase 10.1364/JOSAB.15.000652} {\bibfield
  {journal} {\bibinfo  {journal} {J. Opt. Soc. Am. B}\ }\textbf {\bibinfo
  {volume} {15}},\ \bibinfo {pages} {652} (\bibinfo {year} {1998})}\BibitemShut
  {NoStop}%
\bibitem [{\citenamefont {Borri}\ \emph {et~al.}(2000)\citenamefont {Borri},
  \citenamefont {Langbein}, \citenamefont {Woggon}, \citenamefont {Jensen},\
  and\ \citenamefont {Hvam}}]{Borri2000}%
  \BibitemOpen
  \bibfield  {author} {\bibinfo {author} {\bibfnamefont {P.}~\bibnamefont
  {Borri}}, \bibinfo {author} {\bibfnamefont {W.}~\bibnamefont {Langbein}},
  \bibinfo {author} {\bibfnamefont {U.}~\bibnamefont {Woggon}}, \bibinfo
  {author} {\bibfnamefont {J.~R.}\ \bibnamefont {Jensen}}, \ and\ \bibinfo
  {author} {\bibfnamefont {J.~M.}\ \bibnamefont {Hvam}},\ }\href {\doibase
  10.1103/PhysRevB.62.R7763} {\bibfield  {journal} {\bibinfo  {journal} {Phys.
  Rev. B}\ }\textbf {\bibinfo {volume} {62}},\ \bibinfo {pages} {R7763}
  (\bibinfo {year} {2000})}\BibitemShut {NoStop}%
\bibitem [{\citenamefont {Borri}\ \emph {et~al.}(2003)\citenamefont {Borri},
  \citenamefont {Langbein}, \citenamefont {Woggon}, \citenamefont {Esser},
  \citenamefont {Jensen},\ and\ \citenamefont {rn~M~Hvam}}]{Borri2003}%
  \BibitemOpen
  \bibfield  {author} {\bibinfo {author} {\bibfnamefont {P.}~\bibnamefont
  {Borri}}, \bibinfo {author} {\bibfnamefont {W.}~\bibnamefont {Langbein}},
  \bibinfo {author} {\bibfnamefont {U.}~\bibnamefont {Woggon}}, \bibinfo
  {author} {\bibfnamefont {A.}~\bibnamefont {Esser}}, \bibinfo {author}
  {\bibfnamefont {J.~R.}\ \bibnamefont {Jensen}}, \ and\ \bibinfo {author}
  {\bibfnamefont {J.}~\bibnamefont {rn~M~Hvam}},\ }\href {\doibase
  10.1088/0268-1242/18/10/309} {\bibfield  {journal} {\bibinfo  {journal}
  {Semiconductor Science and Technology}\ }\textbf {\bibinfo {volume} {18}},\
  \bibinfo {pages} {S351} (\bibinfo {year} {2003})}\BibitemShut {NoStop}%
\bibitem [{\citenamefont {Ivanov}\ \emph {et~al.}(2004)\citenamefont {Ivanov},
  \citenamefont {Borri}, \citenamefont {Langbein},\ and\ \citenamefont
  {Woggon}}]{Ivanov2004}%
  \BibitemOpen
  \bibfield  {author} {\bibinfo {author} {\bibfnamefont {A.~L.}\ \bibnamefont
  {Ivanov}}, \bibinfo {author} {\bibfnamefont {P.}~\bibnamefont {Borri}},
  \bibinfo {author} {\bibfnamefont {W.}~\bibnamefont {Langbein}}, \ and\
  \bibinfo {author} {\bibfnamefont {U.}~\bibnamefont {Woggon}},\ }\href
  {\doibase 10.1103/PhysRevB.69.075312} {\bibfield  {journal} {\bibinfo
  {journal} {Phys. Rev. B}\ }\textbf {\bibinfo {volume} {69}},\ \bibinfo
  {pages} {075312} (\bibinfo {year} {2004})}\BibitemShut {NoStop}%
\bibitem [{\citenamefont {Chin}\ \emph {et~al.}(2010)\citenamefont {Chin},
  \citenamefont {Grimm}, \citenamefont {Julienne},\ and\ \citenamefont
  {Tiesinga}}]{Chin2010}%
  \BibitemOpen
  \bibfield  {author} {\bibinfo {author} {\bibfnamefont {C.}~\bibnamefont
  {Chin}}, \bibinfo {author} {\bibfnamefont {R.}~\bibnamefont {Grimm}},
  \bibinfo {author} {\bibfnamefont {P.}~\bibnamefont {Julienne}}, \ and\
  \bibinfo {author} {\bibfnamefont {E.}~\bibnamefont {Tiesinga}},\ }\href
  {\doibase 10.1103/RevModPhys.82.1225} {\bibfield  {journal} {\bibinfo
  {journal} {Rev. Mod. Phys.}\ }\textbf {\bibinfo {volume} {82}},\ \bibinfo
  {pages} {1225} (\bibinfo {year} {2010})}\BibitemShut {NoStop}%
\bibitem [{\citenamefont {Kokkelmans}(2014)}]{Kokkelmans2014}%
  \BibitemOpen
  \bibfield  {author} {\bibinfo {author} {\bibfnamefont {S.}~\bibnamefont
  {Kokkelmans}},\ }in\ \href@noop {} {\emph {\bibinfo {booktitle} {Quantum Gas
  Experiments}}},\ \bibinfo {editor} {edited by\ \bibinfo {editor}
  {\bibfnamefont {P.}~\bibnamefont {T\''{o}rm\''{a}}}\ and\ \bibinfo {editor}
  {\bibfnamefont {K.}~\bibnamefont {Sengstock}}}\ (\bibinfo  {publisher}
  {Imperial College Press},\ \bibinfo {year} {2014})\ pp.\ \bibinfo {pages}
  {63--85}\BibitemShut {NoStop}%
\bibitem [{\citenamefont {Takemura}\ \emph {et~al.}(2016)\citenamefont
  {Takemura}, \citenamefont {Anderson}, \citenamefont {Biswas}, \citenamefont
  {Navadeh-Toupchi}, \citenamefont {Oberli}, \citenamefont {Portella-Oberli},\
  and\ \citenamefont {Deveaud}}]{Takemura2016}%
  \BibitemOpen
  \bibfield  {author} {\bibinfo {author} {\bibfnamefont {N.}~\bibnamefont
  {Takemura}}, \bibinfo {author} {\bibfnamefont {M.~D.}\ \bibnamefont
  {Anderson}}, \bibinfo {author} {\bibfnamefont {S.}~\bibnamefont {Biswas}},
  \bibinfo {author} {\bibfnamefont {M.}~\bibnamefont {Navadeh-Toupchi}},
  \bibinfo {author} {\bibfnamefont {D.~Y.}\ \bibnamefont {Oberli}}, \bibinfo
  {author} {\bibfnamefont {M.~T.}\ \bibnamefont {Portella-Oberli}}, \ and\
  \bibinfo {author} {\bibfnamefont {B.}~\bibnamefont {Deveaud}},\ }\href
  {\doibase 10.1103/PhysRevB.94.195301} {\bibfield  {journal} {\bibinfo
  {journal} {Phys. Rev. B}\ }\textbf {\bibinfo {volume} {94}},\ \bibinfo
  {pages} {195301} (\bibinfo {year} {2016})}\BibitemShut {NoStop}%
\bibitem [{\citenamefont {Ciuti}\ \emph {et~al.}(2003)\citenamefont {Ciuti},
  \citenamefont {Schwendimann},\ and\ \citenamefont {Quattropani}}]{Ciuti2003}%
  \BibitemOpen
  \bibfield  {author} {\bibinfo {author} {\bibfnamefont {C.}~\bibnamefont
  {Ciuti}}, \bibinfo {author} {\bibfnamefont {P.}~\bibnamefont {Schwendimann}},
  \ and\ \bibinfo {author} {\bibfnamefont {A.}~\bibnamefont {Quattropani}},\
  }\href {\doibase 10.1088/0268-1242/18/10/301} {\bibfield  {journal} {\bibinfo
   {journal} {Semiconductor Science and Technology}\ }\textbf {\bibinfo
  {volume} {18}},\ \bibinfo {pages} {S279} (\bibinfo {year}
  {2003})}\BibitemShut {NoStop}%
\bibitem [{\citenamefont {Combescot}\ \emph {et~al.}(2007)\citenamefont
  {Combescot}, \citenamefont {Dupertuis},\ and\ \citenamefont
  {Betbeder-Matibet}}]{Combescot2007}%
  \BibitemOpen
  \bibfield  {author} {\bibinfo {author} {\bibfnamefont {M.}~\bibnamefont
  {Combescot}}, \bibinfo {author} {\bibfnamefont {M.~A.}\ \bibnamefont
  {Dupertuis}}, \ and\ \bibinfo {author} {\bibfnamefont {O.}~\bibnamefont
  {Betbeder-Matibet}},\ }\href {\doibase 10.1209/0295-5075/79/17001} {\bibfield
   {journal} {\bibinfo  {journal} {Europhysics Letters ({EPL})}\ }\textbf
  {\bibinfo {volume} {79}},\ \bibinfo {pages} {17001} (\bibinfo {year}
  {2007})}\BibitemShut {NoStop}%
\bibitem [{\citenamefont {Ciuti}\ and\ \citenamefont
  {Carusotto}(2005)}]{Ciuti2005}%
  \BibitemOpen
  \bibfield  {author} {\bibinfo {author} {\bibfnamefont {C.}~\bibnamefont
  {Ciuti}}\ and\ \bibinfo {author} {\bibfnamefont {I.}~\bibnamefont
  {Carusotto}},\ }\href {\doibase 10.1002/pssb.200560961} {\bibfield  {journal}
  {\bibinfo  {journal} {physica status solidi (b)}\ }\textbf {\bibinfo {volume}
  {242}},\ \bibinfo {pages} {2224} (\bibinfo {year} {2005})}\BibitemShut
  {NoStop}%
\bibitem [{\citenamefont {Massignan}\ \emph {et~al.}(2014)\citenamefont
  {Massignan}, \citenamefont {Zaccanti},\ and\ \citenamefont
  {Bruun}}]{Massignan2014}%
  \BibitemOpen
  \bibfield  {author} {\bibinfo {author} {\bibfnamefont {P.}~\bibnamefont
  {Massignan}}, \bibinfo {author} {\bibfnamefont {M.}~\bibnamefont {Zaccanti}},
  \ and\ \bibinfo {author} {\bibfnamefont {G.~M.}\ \bibnamefont {Bruun}},\
  }\href {http://stacks.iop.org/0034-4885/77/i=3/a=034401} {\bibfield
  {journal} {\bibinfo  {journal} {Rep. Progr. Phys.}\ }\textbf {\bibinfo
  {volume} {77}},\ \bibinfo {pages} {034401} (\bibinfo {year}
  {2014})}\BibitemShut {NoStop}%
\bibitem [{\citenamefont {Rath}\ and\ \citenamefont
  {Schmidt}(2013)}]{Rath2013}%
  \BibitemOpen
  \bibfield  {author} {\bibinfo {author} {\bibfnamefont {S.~P.}\ \bibnamefont
  {Rath}}\ and\ \bibinfo {author} {\bibfnamefont {R.}~\bibnamefont {Schmidt}},\
  }\href {\doibase 10.1103/PhysRevA.88.053632} {\bibfield  {journal} {\bibinfo
  {journal} {Phys. Rev. A}\ }\textbf {\bibinfo {volume} {88}},\ \bibinfo
  {pages} {053632} (\bibinfo {year} {2013})}\BibitemShut {NoStop}%
\bibitem [{\citenamefont {Pe\~na Ardila}\ \emph {et~al.}(2019)\citenamefont
  {Pe\~na Ardila}, \citenamefont {J\o{}rgensen}, \citenamefont {Pohl},
  \citenamefont {Giorgini}, \citenamefont {Bruun},\ and\ \citenamefont
  {Arlt}}]{Ardila2019}%
  \BibitemOpen
  \bibfield  {author} {\bibinfo {author} {\bibfnamefont {L.~A.}\ \bibnamefont
  {Pe\~na Ardila}}, \bibinfo {author} {\bibfnamefont {N.~B.}\ \bibnamefont
  {J\o{}rgensen}}, \bibinfo {author} {\bibfnamefont {T.}~\bibnamefont {Pohl}},
  \bibinfo {author} {\bibfnamefont {S.}~\bibnamefont {Giorgini}}, \bibinfo
  {author} {\bibfnamefont {G.~M.}\ \bibnamefont {Bruun}}, \ and\ \bibinfo
  {author} {\bibfnamefont {J.~J.}\ \bibnamefont {Arlt}},\ }\href {\doibase
  10.1103/PhysRevA.99.063607} {\bibfield  {journal} {\bibinfo  {journal} {Phys.
  Rev. A}\ }\textbf {\bibinfo {volume} {99}},\ \bibinfo {pages} {063607}
  (\bibinfo {year} {2019})}\BibitemShut {NoStop}%
\bibitem [{\citenamefont {Randeria}\ \emph {et~al.}(1990)\citenamefont
  {Randeria}, \citenamefont {Duan},\ and\ \citenamefont
  {Shieh}}]{Randeria1990}%
  \BibitemOpen
  \bibfield  {author} {\bibinfo {author} {\bibfnamefont {M.}~\bibnamefont
  {Randeria}}, \bibinfo {author} {\bibfnamefont {J.-M.}\ \bibnamefont {Duan}},
  \ and\ \bibinfo {author} {\bibfnamefont {L.-Y.}\ \bibnamefont {Shieh}},\
  }\href {\doibase 10.1103/PhysRevB.41.327} {\bibfield  {journal} {\bibinfo
  {journal} {Phys. Rev. B}\ }\textbf {\bibinfo {volume} {41}},\ \bibinfo
  {pages} {327} (\bibinfo {year} {1990})}\BibitemShut {NoStop}%
\bibitem [{\citenamefont {Wouters}(2007)}]{Wouters2007}%
  \BibitemOpen
  \bibfield  {author} {\bibinfo {author} {\bibfnamefont {M.}~\bibnamefont
  {Wouters}},\ }\href {\doibase 10.1103/PhysRevB.76.045319} {\bibfield
  {journal} {\bibinfo  {journal} {Phys. Rev. B}\ }\textbf {\bibinfo {volume}
  {76}},\ \bibinfo {pages} {045319} (\bibinfo {year} {2007})}\BibitemShut
  {NoStop}%
\bibitem [{\citenamefont {Carusotto}\ \emph {et~al.}(2010)\citenamefont
  {Carusotto}, \citenamefont {Volz},\ and\ \citenamefont
  {Imamo{\u{g}}lu}}]{Carusotto2010}%
  \BibitemOpen
  \bibfield  {author} {\bibinfo {author} {\bibfnamefont {I.}~\bibnamefont
  {Carusotto}}, \bibinfo {author} {\bibfnamefont {T.}~\bibnamefont {Volz}}, \
  and\ \bibinfo {author} {\bibfnamefont {A.}~\bibnamefont {Imamo{\u{g}}lu}},\
  }\href {\doibase 10.1209/0295-5075/90/37001} {\bibfield  {journal} {\bibinfo
  {journal} {{EPL} (Europhysics Letters)}\ }\textbf {\bibinfo {volume} {90}},\
  \bibinfo {pages} {37001} (\bibinfo {year} {2010})}\BibitemShut {NoStop}%
\bibitem [{\citenamefont {{Tan}}\ \emph {et~al.}(2019)\citenamefont {{Tan}},
  \citenamefont {{Cotlet}}, \citenamefont {{Bergschneider}}, \citenamefont
  {{Schmidt}}, \citenamefont {{Back}}, \citenamefont {{Shimazaki}},
  \citenamefont {{Kroner}},\ and\ \citenamefont {{Imamoglu}}}]{Tan2019}%
  \BibitemOpen
  \bibfield  {author} {\bibinfo {author} {\bibfnamefont {L.~B.}\ \bibnamefont
  {{Tan}}}, \bibinfo {author} {\bibfnamefont {O.}~\bibnamefont {{Cotlet}}},
  \bibinfo {author} {\bibfnamefont {A.}~\bibnamefont {{Bergschneider}}},
  \bibinfo {author} {\bibfnamefont {R.}~\bibnamefont {{Schmidt}}}, \bibinfo
  {author} {\bibfnamefont {P.}~\bibnamefont {{Back}}}, \bibinfo {author}
  {\bibfnamefont {Y.}~\bibnamefont {{Shimazaki}}}, \bibinfo {author}
  {\bibfnamefont {M.}~\bibnamefont {{Kroner}}}, \ and\ \bibinfo {author}
  {\bibfnamefont {A.}~\bibnamefont {{Imamoglu}}},\ }\href@noop {} {\bibfield
  {journal} {\bibinfo  {journal} {arXiv e-prints}\ ,\ \bibinfo {eid}
  {arXiv:1903.05640}} (\bibinfo {year} {2019})},\ \Eprint
  {http://arxiv.org/abs/1903.05640} {arXiv:1903.05640 [cond-mat.mes-hall]}
  \BibitemShut {NoStop}%
\bibitem [{\citenamefont {Fetter}\ and\ \citenamefont
  {Walecka}(1971)}]{Fetter1971}%
  \BibitemOpen
  \bibfield  {author} {\bibinfo {author} {\bibfnamefont {A.}~\bibnamefont
  {Fetter}}\ and\ \bibinfo {author} {\bibfnamefont {J.}~\bibnamefont
  {Walecka}},\ }\href@noop {} {\emph {\bibinfo {title} {Quantum Theory of
  Many-Particle Systems}}},\ Dover Books on Physics Series\ (\bibinfo
  {publisher} {Dover Publications},\ \bibinfo {year} {1971})\BibitemShut
  {NoStop}%
\bibitem [{\citenamefont {Grassi~Alessi}\ \emph {et~al.}(2000)\citenamefont
  {Grassi~Alessi}, \citenamefont {Fragano}, \citenamefont {Patan\`e},
  \citenamefont {Capizzi}, \citenamefont {Runge},\ and\ \citenamefont
  {Zimmermann}}]{Alessi2000}%
  \BibitemOpen
  \bibfield  {author} {\bibinfo {author} {\bibfnamefont {M.}~\bibnamefont
  {Grassi~Alessi}}, \bibinfo {author} {\bibfnamefont {F.}~\bibnamefont
  {Fragano}}, \bibinfo {author} {\bibfnamefont {A.}~\bibnamefont {Patan\`e}},
  \bibinfo {author} {\bibfnamefont {M.}~\bibnamefont {Capizzi}}, \bibinfo
  {author} {\bibfnamefont {E.}~\bibnamefont {Runge}}, \ and\ \bibinfo {author}
  {\bibfnamefont {R.}~\bibnamefont {Zimmermann}},\ }\href {\doibase
  10.1103/PhysRevB.61.10985} {\bibfield  {journal} {\bibinfo  {journal} {Phys.
  Rev. B}\ }\textbf {\bibinfo {volume} {61}},\ \bibinfo {pages} {10985}
  (\bibinfo {year} {2000})}\BibitemShut {NoStop}%
\bibitem [{\citenamefont {Yamamoto}\ \emph {et~al.}(2000)\citenamefont
  {Yamamoto}, \citenamefont {Tassone},\ and\ \citenamefont
  {Cao}}]{Yamamoto2000}%
  \BibitemOpen
  \bibfield  {author} {\bibinfo {author} {\bibfnamefont {Y.}~\bibnamefont
  {Yamamoto}}, \bibinfo {author} {\bibfnamefont {T.}~\bibnamefont {Tassone}}, \
  and\ \bibinfo {author} {\bibfnamefont {H.}~\bibnamefont {Cao}},\ }\href@noop
  {} {\emph {\bibinfo {title} {Semiconductor Cavity Quantum
  Electrodynamics}}},\ Springer Tracts in Modern Physics\ (\bibinfo
  {publisher} {Springer-Verlag},\ \bibinfo {year} {2000})\BibitemShut {NoStop}%
\bibitem [{\citenamefont {Deveaud-Pl\'edran}\ and\ \citenamefont
  {Lagoudakis}(2012)}]{Deveaud2012}%
  \BibitemOpen
  \bibfield  {author} {\bibinfo {author} {\bibfnamefont {B.}~\bibnamefont
  {Deveaud-Pl\'edran}}\ and\ \bibinfo {author} {\bibfnamefont {K.~G.}\
  \bibnamefont {Lagoudakis}},\ }in\ \href@noop {} {\emph {\bibinfo {booktitle}
  {Exciton Polaritons in Microcavities. New Frontiers}}},\ \bibinfo {series and
  number} {Springer Series in Solid-State Sciences},\ \bibinfo {editor} {edited
  by\ \bibinfo {editor} {\bibfnamefont {D.}~\bibnamefont {Sanvitto}}\ and\
  \bibinfo {editor} {\bibfnamefont {V.}~\bibnamefont {Timofeev}}}\ (\bibinfo
  {publisher} {Springer-Verlag},\ \bibinfo {year} {2012})\ pp.\ \bibinfo
  {pages} {233--243}\BibitemShut {NoStop}%
\bibitem [{Note1()}]{Note1}%
  \BibitemOpen
  \bibinfo {note} {The reported value of the density $n_{pu}$ in $\unhbox
  \voidb@x \hbox {cm}^{-2}$ depends on the chosen value of the exciton mass. As
  typical values range from $m_{x}=0.1m_{e}$ to $0.25m_{e}$ we could have an
  uncertainty of 50\% of the chosen value for the density.}\BibitemShut {Stop}%
\bibitem [{\citenamefont {Birkedal}\ \emph {et~al.}(1996)\citenamefont
  {Birkedal}, \citenamefont {Singh}, \citenamefont {Lyssenko}, \citenamefont
  {Erland},\ and\ \citenamefont {Hvam}}]{Birkedal1996}%
  \BibitemOpen
  \bibfield  {author} {\bibinfo {author} {\bibfnamefont {D.}~\bibnamefont
  {Birkedal}}, \bibinfo {author} {\bibfnamefont {J.}~\bibnamefont {Singh}},
  \bibinfo {author} {\bibfnamefont {V.~G.}\ \bibnamefont {Lyssenko}}, \bibinfo
  {author} {\bibfnamefont {J.}~\bibnamefont {Erland}}, \ and\ \bibinfo {author}
  {\bibfnamefont {J.~M.}\ \bibnamefont {Hvam}},\ }\href {\doibase
  10.1103/PhysRevLett.76.672} {\bibfield  {journal} {\bibinfo  {journal} {Phys.
  Rev. Lett.}\ }\textbf {\bibinfo {volume} {76}},\ \bibinfo {pages} {672}
  (\bibinfo {year} {1996})}\BibitemShut {NoStop}%
\bibitem [{\citenamefont {{Levinsen}}\ \emph {et~al.}(2018)\citenamefont
  {{Levinsen}}, \citenamefont {{Marchetti}}, \citenamefont {{Keeling}},\ and\
  \citenamefont {{Parish}}}]{Levinsen2018}%
  \BibitemOpen
  \bibfield  {author} {\bibinfo {author} {\bibfnamefont {J.}~\bibnamefont
  {{Levinsen}}}, \bibinfo {author} {\bibfnamefont {F.~M.}\ \bibnamefont
  {{Marchetti}}}, \bibinfo {author} {\bibfnamefont {J.}~\bibnamefont
  {{Keeling}}}, \ and\ \bibinfo {author} {\bibfnamefont {M.~M.}\ \bibnamefont
  {{Parish}}},\ }\href@noop {} {\bibfield  {journal} {\bibinfo  {journal}
  {arXiv e-prints}\ ,\ \bibinfo {eid} {arXiv:1806.10835}} (\bibinfo {year}
  {2018})},\ \Eprint {http://arxiv.org/abs/1806.10835} {arXiv:1806.10835
  [cond-mat.quant-gas]} \BibitemShut {NoStop}%
\bibitem [{\citenamefont {Bao}\ \emph {et~al.}(2018)\citenamefont {Bao},
  \citenamefont {Liu}, \citenamefont {Xue}, \citenamefont {Zheng},
  \citenamefont {Tao}, \citenamefont {Wang}, \citenamefont {Xia}, \citenamefont
  {Zhao}, \citenamefont {Kim}, \citenamefont {Yang}, \citenamefont {Li},
  \citenamefont {Wang}, \citenamefont {Wang}, \citenamefont {Wang},
  \citenamefont {MacDonald},\ and\ \citenamefont {Zhang}}]{bao2018}%
  \BibitemOpen
  \bibfield  {author} {\bibinfo {author} {\bibfnamefont {W.}~\bibnamefont
  {Bao}}, \bibinfo {author} {\bibfnamefont {X.}~\bibnamefont {Liu}}, \bibinfo
  {author} {\bibfnamefont {F.}~\bibnamefont {Xue}}, \bibinfo {author}
  {\bibfnamefont {F.}~\bibnamefont {Zheng}}, \bibinfo {author} {\bibfnamefont
  {R.}~\bibnamefont {Tao}}, \bibinfo {author} {\bibfnamefont {S.}~\bibnamefont
  {Wang}}, \bibinfo {author} {\bibfnamefont {Y.}~\bibnamefont {Xia}}, \bibinfo
  {author} {\bibfnamefont {M.}~\bibnamefont {Zhao}}, \bibinfo {author}
  {\bibfnamefont {J.}~\bibnamefont {Kim}}, \bibinfo {author} {\bibfnamefont
  {S.}~\bibnamefont {Yang}}, \bibinfo {author} {\bibfnamefont {Q.}~\bibnamefont
  {Li}}, \bibinfo {author} {\bibfnamefont {Y.}~\bibnamefont {Wang}}, \bibinfo
  {author} {\bibfnamefont {Y.}~\bibnamefont {Wang}}, \bibinfo {author}
  {\bibfnamefont {L.-W.}\ \bibnamefont {Wang}}, \bibinfo {author}
  {\bibfnamefont {A.}~\bibnamefont {MacDonald}}, \ and\ \bibinfo {author}
  {\bibfnamefont {X.}~\bibnamefont {Zhang}},\ }\href@noop {} {\enquote
  {\bibinfo {title} {Observation of rydberg exciton polaritons and their
  condensate in a perovskite cavity},}\ } (\bibinfo {year} {2018}),\ \Eprint
  {http://arxiv.org/abs/1803.07282} {arXiv:1803.07282 [cond-mat.mtrl-sci]}
  \BibitemShut {NoStop}%
\bibitem [{\citenamefont {Rodriguez}\ \emph {et~al.}(2016)\citenamefont
  {Rodriguez}, \citenamefont {Amo}, \citenamefont {Sagnes}, \citenamefont
  {Le~Gratiet}, \citenamefont {Galopin}, \citenamefont {Lema\^{i}tre},\ and\
  \citenamefont {Bloch}}]{Rodriguez2016}%
  \BibitemOpen
  \bibfield  {author} {\bibinfo {author} {\bibfnamefont {S.~R.~K.}\
  \bibnamefont {Rodriguez}}, \bibinfo {author} {\bibfnamefont {A.}~\bibnamefont
  {Amo}}, \bibinfo {author} {\bibfnamefont {I.}~\bibnamefont {Sagnes}},
  \bibinfo {author} {\bibfnamefont {L.}~\bibnamefont {Le~Gratiet}}, \bibinfo
  {author} {\bibfnamefont {E.}~\bibnamefont {Galopin}}, \bibinfo {author}
  {\bibfnamefont {A.}~\bibnamefont {Lema\^{i}tre}}, \ and\ \bibinfo {author}
  {\bibfnamefont {J.}~\bibnamefont {Bloch}},\ }\href
  {https://doi.org/10.1038/ncomms11887} {\bibfield  {journal} {\bibinfo
  {journal} {Nature Communications}\ }\textbf {\bibinfo {volume} {7}},\
  \bibinfo {pages} {11887 EP } (\bibinfo {year} {2016})},\ \bibinfo {note}
  {article}\BibitemShut {NoStop}%
\bibitem [{\citenamefont {Sidler}\ \emph {et~al.}(2016)\citenamefont {Sidler},
  \citenamefont {Back}, \citenamefont {Cotlet}, \citenamefont {Srivastava},
  \citenamefont {Fink}, \citenamefont {Kroner}, \citenamefont {Demler},\ and\
  \citenamefont {Imamoglu}}]{Sidler2016}%
  \BibitemOpen
  \bibfield  {author} {\bibinfo {author} {\bibfnamefont {M.}~\bibnamefont
  {Sidler}}, \bibinfo {author} {\bibfnamefont {P.}~\bibnamefont {Back}},
  \bibinfo {author} {\bibfnamefont {O.}~\bibnamefont {Cotlet}}, \bibinfo
  {author} {\bibfnamefont {A.}~\bibnamefont {Srivastava}}, \bibinfo {author}
  {\bibfnamefont {T.}~\bibnamefont {Fink}}, \bibinfo {author} {\bibfnamefont
  {M.}~\bibnamefont {Kroner}}, \bibinfo {author} {\bibfnamefont
  {E.}~\bibnamefont {Demler}}, \ and\ \bibinfo {author} {\bibfnamefont
  {A.}~\bibnamefont {Imamoglu}},\ }\href {https://doi.org/10.1038/nphys3949}
  {\bibfield  {journal} {\bibinfo  {journal} {Nature Physics}\ }\textbf
  {\bibinfo {volume} {13}},\ \bibinfo {pages} {255 EP } (\bibinfo {year}
  {2016})}\BibitemShut {NoStop}%
\end{thebibliography}%
%%%%%%%%%%%%%%%%%%%%%%%%%%%%%%%%%%%%%%%%%%%%%%%%%%

%%%%%%%%%%%%%%%%%%%%%%%%%%%%%%%%%%%%%%%%%%%%%%%%%%
%%%%%%%%%%%%%%%%%%%%%%%%%%%%%%%%%%%%%%%%%%%%%%%%%%

\end{document}